\documentclass[aps, twocolumn, groupedaddress, nofootinbib, preprintnumbers]{revtex4}
\usepackage{float}
\usepackage[dvips]{graphicx}
\usepackage{color}
\usepackage{amsmath,amssymb,slashed}
\usepackage{tikz}
\usepackage{hyperref}
\usepackage{soul}
\usepackage{tabularx}
\newcommand{\be}{\begin{equation}}
\newcommand{\ee}{\end{equation}}

\graphicspath{{images/}}

\begin{document}
	
\preprint{YITP-SB-2020-38}
\title{Scale-dependent halo bias and the squeezed limit bispectrum in the presence of radiation}
\author{Charuhas Shiveshwarkar, Drew Jamieson, and Marilena Loverde \\{\it{\small  C.N. Yang Institute for Theoretical Physics, Department of Physics \& Astronomy, Stony Brook University, Stony Brook, NY 11794}}}

\begin{abstract}
We investigate the gravitational effect of large-scale radiation perturbations on small-scale structure formation. In addition to making the growth of matter perturbations scale dependent, the free-streaming of radiation also affects the coupling between structure formation at small and large scales. We study this using Separate Universe N-body simulations to compute the (isotropized) squeezed-limit matter bispectrum and the linear halo bias. Our results show that the scale dependence in the growth of long-wavelength matter perturbations, caused by radiation, translates into these quantities acquiring a non-trivial scale-dependence at $k\lesssim 0.05$ Mpc$^{-1}$. In a universe with radiation composed of cosmic microwave background photons and three species of massless neutrinos, the bias of halos with $b = 2$ at high $k$ will decrease by $0.29\%,\ 0.45\%$ and  $0.8\%$ between $k  =  0.05$ Mpc$^{-1}$ and $k = 0.0005$ Mpc$^{-1}$ at redshifts $z=0,\ 1$, and $3$ respectively. For objects with $b\gg1$, these differences approach $0.43\%,\ 0.68\%$ and $1.2\%$ respectively.
\end{abstract}

\maketitle

\section{Introduction}
Along with the cosmic microwave background (CMB), observations of large-scale structure in the Universe are an important tool for obtaining constraints on cosmological parameters. The key observables therein include correlation functions of tracers of the matter density fluctuation, such as galaxies, quasars, or the intensity of emission and absorption lines from intergalactic gas. Correlation functions of these tracers, for instance their power spectrum and bispectrum, produce biased estimates of correlations of the matter density field. At large scales, however, there is a linear relationship between fluctuations in the tracer population and fluctuations in the underlying matter density field. This proportionality factor relating the two is called the {\em bias} of the tracer (see, e.g. \cite{Desjacques:2016bnm}, for a review). Accurately modeling these observables requires precise predictions for both the bias of tracers and the gravitational evolution of the matter density field, two non-trivial challenges arising from the non-linear nature of structure formation.

Density fluctuations in the early universe evolved through gravitational collapse into structures like the galaxies and the galaxy clusters observable today. In a universe with only a cosmological constant $\Lambda$ and cold dark matter (CDM), the growth of matter density fluctuations in the linear regime is scale-independent (for a review, see \cite{Bernardeau:2001qr}). The presence of free-streaming components, such as CMB photons and neutrinos, introduces a new scale into the dynamics of structure formation: namely their respective free-streaming scales. A free-streaming scale acts as a Jeans scale in the sense that a free-streaming component does not cluster with CDM below this scale. This causes the growth of matter density fluctuations at scales smaller than the free-streaming scale to be suppressed in comparison to the growth at larger scales  (e.g. \cite{lesgourgues2006massive}). This is a well-known effect on the linear growth of matter density fluctuations which, as we show, modifies the coupling between structure formation at small and large scales non-trivially. Specifically, we will show that free-streaming introduces a scale-dependence to the halo bias and the squeezed-limit matter bispectrum.  It is important to note that while the free-streaming scale for massless particles is the Hubble scale ($c/aH$), the effects we study persist down to observable scales ($k \sim 5\times 10^{-2}~\textrm{Mpc}^{-1}$). The new scale-dependent changes are small ($\lesssim 1\%$ on the halo bias for $z\lesssim 3$), but potentially important for future measurements that demand exquisite control over systematics for galaxy clustering observations on large scales (for example, those targeting primordial non-Gaussianity with $f_{NL}\sim 1$ \cite{Dore:2014cca,Meerburg:2019qqi}). Moreover, these changes to nonlinear structure are a real gravitational effect of radiation at late times that should nonetheless be quantified. 

In this paper, we use the Separate Universe technique to compute responses of small-scale observables to large-scale background matter and radiation density fluctuations. A long-wavelength matter density perturbation affects the cosmic expansion history locally as would be observed by a small-scale observer situated within it. The long-wavelength density perturbation can thus be absorbed into the background cosmology to yield a set of \textit{local} cosmological parameters governing the growth of small-scale structure within the long-wavelength density perturbation. Performing N-body simulations within this \textit{Separate Universe} gives a way to obtain the responses of small-scale observables to the background density perturbation well beyond the linear level. This technique has been used to determine halo and void bias, the position-dependent power spectrum, and position-dependent one-point statistics in $\Lambda$CDM cosmologies \cite{McDonald:2001fe, Sirko:2005uz, Gnedin:2011kj, Wagner:2014aka, Li:2014sga, Li:2014jra, Li:2015jsz, Chiang:2014oga, Manzotti:2014wca, Baldauf:2015vio, 
Lazeyras:2015lgp, Paranjape:2016pbh, Chan:2019yzq, Jamieson:2019dmp, Jamieson:2020wxf}. A framework to extend the Separate Universe technique beyond $\Lambda$CDM was presented in \cite{Hu:2016ssz} and applied to cosmologies with massive neutrinos \cite{Chiang:2017vuk}, dynamical dark energy \cite{ Chiang:2016vxa, Jamieson:2018biz}, and isocurvature modes \cite{Jamieson:2018biz, Barreira:2019qdl}. 

\begin{table}
\centering
\begin{tabular}{||c|c|c|c|c|c|c||} 
 \hline
 $ T_{cmb}$ & $\Omega_{cdm}$ & $\Omega_{b}$ & $\sigma_8$ & $n_s$ & YHe  & $h$\\ [0.5ex] 
 \hline\hline
 2.725 K & 0.25 & 0.05 & 0.83 & 0.95 & 0.24 & 0.7 \\ [1ex]
 \hline
\end{tabular}
\caption{Parameters of the background cosmology. We assumed the Universe is flat and when we varied the energy density in radiation, we adjusted the value of $\Omega_\Lambda$ to satisfy $\Omega_\Lambda = 1 - \Omega_{cdm} - \Omega_b - \Omega_{rad}$.}
\label{cosmology}
\end{table}

Throughout this paper, we work in a flat $\nu\Lambda$CDM cosmology with cosmological parameters as given in Table \ref{cosmology}. We will vary the energy density in radiation and accordingly adjust the value of $\Omega_\Lambda$ to satisfy  $\Omega_\Lambda = 1 - \Omega_{cdm} - \Omega_b - \Omega_{rad}$. Note that we specify the helium abundance separately, rather than adjusting for consistency with big bang nucleosynthesis \cite{Pisanti:2007hk}, because we will consider examples with very large deviations from the radiation density in the standard cosmology. For the purposes of this work, we regard baryons and CDM together as a single fluid with $ \Omega_{c} = 0.30$. Henceforth ``CDM" or matter with subscript $\rho_c$ refers to both CDM and baryons. The linear evolution of density perturbations is obtained using the public code CLASS \cite{blas2011cosmic}. 

We begin with a review of the Separate Universe formalism in Sec. \ref{sec:Separate Universesetup}. While our goal is to study the effects of large-scale radiation perturbations, we first present calculations of the effects due to neutrinos with degenerate masses in Sec. \ref{sec:analyticalresults}. We vary both the number of species and the mass per species, demonstrating non-trivial scale-dependent effects in the limit $m_\nu \rightarrow 0$. We also demonstrate that the effects are insensitive to the composition of the radiation by comparing results for massless neutrinos to photons, while holding $\Omega_{rad}$ fixed. For our N-body simulations, presented in Sec. \ref{sec:Nbody}, we chose a cosmology with 28 massless neutrinos, corresponding to a radiation density today of $\Omega_{rad} = 3.711\times 10^{-4}$, which is about an order of magnitude larger than the radiation density in our Universe. This large value of $\Omega_{rad} $ is chosen so that we can reliably measure the effects of radiation on the power spectrum response and halo bias with a reasonable number of simulations. We present our results obtained from Separate Universe N-body simulations in Sec. \ref{sec:results from simulations}. In Sec. \ref{sec:bias models} we show that the scale-dependence of the Lagrangian response bias obtained from N-body simulations can be accurately modeled using the power spectrum response computed using one-loop perturbation theory. In Sec. \ref{sec:effects on observables} we discuss the scale dependence and redshift dependence of the Eulerian bias computed using this model in a cosmology similar to our own, with three massless neutrinos. 

\section{Construction of the Separate Universe}
\label{sec:Separate Universesetup}

An observer within a large-scale perturbation to the matter density, $\delta_c$, will observe small-scale density perturbations against a background matter density which includes $\delta_c$ in addition to the unperturbed matter density $\overline{\rho}_{c}$ in the global universe,
\be
 \overline{\rho}_{cW} = \overline{\rho}_{c}(1+\delta_c)\,.
 \ee
The long-wavelength perturbation $\delta_c$ can be absorbed into the background cosmic expansion to yield a set of local cosmological parameters. Using the synchronous gauge for $\delta_c$ and choosing the CDM frame to define the local scale factor gives \cite{Hu:2016ssz}, 
\be
a_W = a \left(1+\delta_c(k_L,a)\right)^{-\frac{1}{3}} \approx a\left[1-\frac{\delta_c(k_L,a)}{3}\right]\,,
\ee
where $k_L$ is the wave number of the perturbation. The local Hubble rate is then,
\be
H_{W} = H\left[1-\frac{1}{3}\delta_c{'}(k_L,a)\right]\,,
\ee
where $' = \frac{d}{d\log a}$. The Separate Universe by construction is dependent on wave number $k_L$ of the large-scale background density perturbation. Scale-dependent evolution of $\delta_c$ leads to scale-dependent Separate Universe responses of the small-scale observables. 

\begin{figure}
  \centering
  \includegraphics[width=\columnwidth]{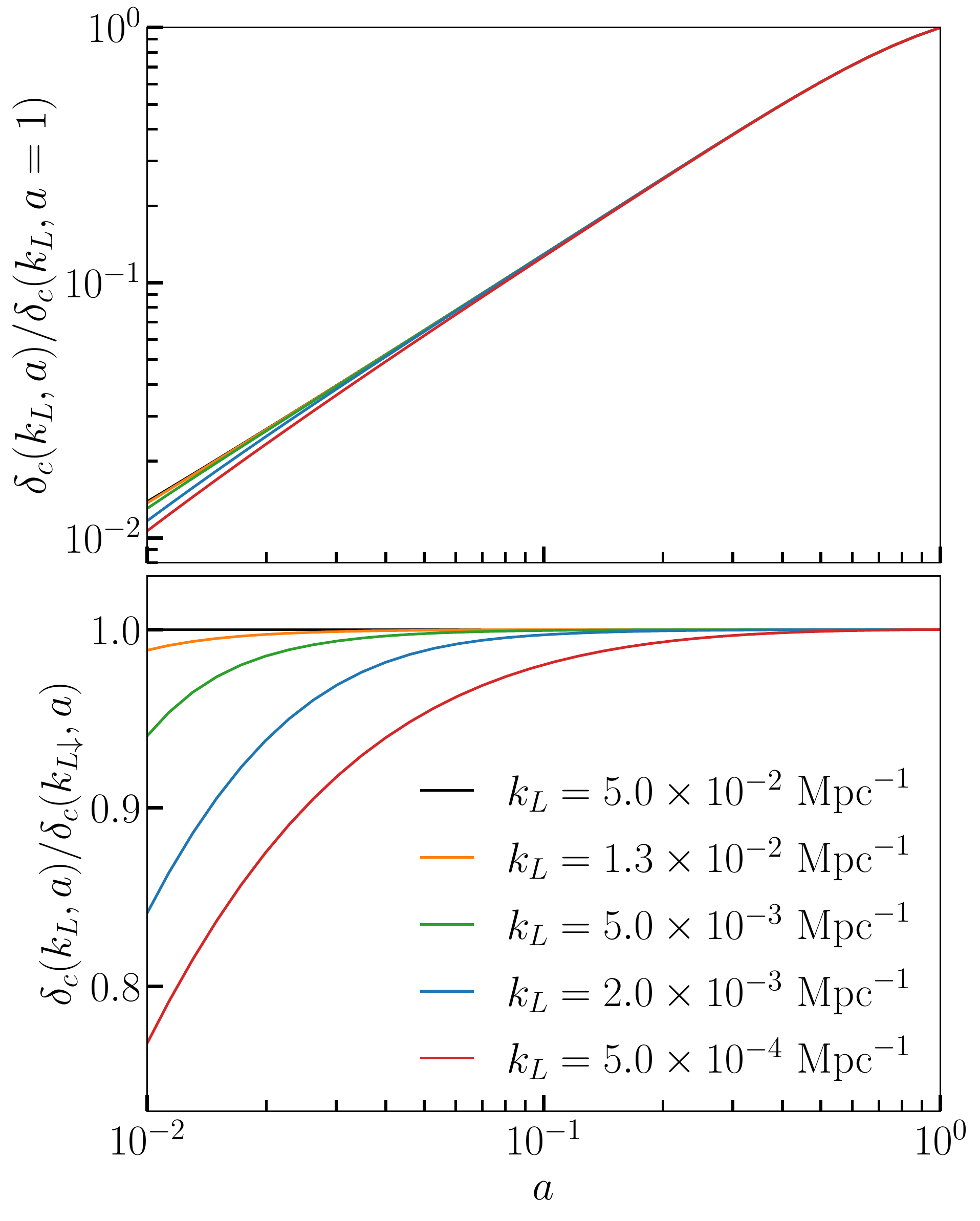}
\caption{(Top) Evolution of $\delta_c(k_L,a)/\delta_c(k_L,a=1)$ for different values of $k_L$.
(Bottom) Evolution of $\delta_c(k_L,a)/\delta_c(k_{L\downarrow} = 5 \times 10^{-2}\ \mathrm{Mpc}^{-1},a)$ for different values of $k_L$. In this figure, the effects of radiation have been enhanced by setting $\Omega_{\mathrm{rad}} = 3.7\times 10^{-4}$.}
\label{fig:deltac}
\end{figure}

The growth of matter density perturbations is suppressed on scales below the Hubble scale due to the free streaming of massless neutrinos and CMB photons \cite{lesgourgues2006massive}. This scale-dependent growth is illustrated in Figure \ref{fig:deltac}, which plots the evolution of $\delta_c(a,k)$ at different large scales, normalized to $\delta_{c0} = \delta_c(a=1)$. Modes with longer wavelengths, which include coherent perturbations in CDM, photons, and massless neutrinos, grow faster than modes with shorter wavelengths due to the free streaming of neutrinos and photons on subhorizon scales. 

\subsection{Linear Growth Factor Response}
\label{ssec:growthresponse}

We are interested in quantifying how small-scale observables are modified by the presence of long-wavelength modes $\delta_c(a, k_L)$ that span the range of scales shown in Fig. \ref{fig:deltac}. We will refer to the changes in small-scale observables, due to the presence of long-wavelength modes as Separate Universe responses. As has been demonstrated in previous works \cite{Chiang:2017vuk, Chiang:2016vxa, Jamieson:2018biz, Jamieson:2019dmp}, much about the Separate Universe responses can be understood by studying how the evolution of linear matter perturbations is modified in the Separate Universe. 

By construction, only the CDM (and baryon) component clusters on scales $k \gg aH$. The linear growth factor of small-scale CDM density perturbations in the Separate Universe, denoted $D_{W}$, is given by the usual linear growth equation with the scale factor, Hubble rate, and matter density replaced by those in the Separate Universe cosmology,
\begin{align}
\label{eq:DW}
\frac{d^2D_{W}}{d\log a_W^2}+\left(2+\frac{d\log H_{W}}{d\log a_W}\right)\frac{dD_{W}}{d\log a_W} = \frac{3}{2}\frac{\Omega_{cW}H_{0W}^2}{H_W^2 a_W^3}D_W\, , 
\end{align}
where $ \Omega_{cW}H_{0W}^2 = \Omega_c H_0^2 $. The Separate Universe linear growth factor can be expanded around the global universe linear growth factor, $D$. To linear order in $\delta_c(k_L,a)$, we denote this as $D_W = D + \epsilon(k_L)$, where 
\begin{eqnarray} \label{eq:Depsilon}
D''+\left(2+\frac{H'}{H}\right)D'-\frac{3}{2}\frac{\Omega_c H_0^2}{H^2 a^3}D &=& 0\,,\\
 \label{eq:epsilon}
\epsilon''+\left(2+\frac{H'}{H}\right)\epsilon'-\frac{3}{2}\frac{\Omega_c H_0^2}{H^2 a^3}\epsilon &=& \frac{2}{3}\delta_{c}'D' \\ &&+\frac{3}{2}\frac{\Omega_c H_0^2}{H^2 a^3}\delta_cD \,.\nonumber
\end{eqnarray} 

To solve these equations, we need to impose appropriate initial conditions for $D_W$ and $\epsilon$. Since we want results that are accurate at $\mathcal{O}(\Omega_{rad})$, including for larger values of $\Omega_{rad}$ than in our own Universe, we impose initial conditions in the radiation dominated era ($a_{i}\approx 10^{-6}$). In this case, the global universe linear growth factor is,
\be
D = C_{1}\log\left(\frac{a}{a_{H}}\right)\,,
\ee 
where $a_{H} \ll a_{i}$ is the scale factor at horizon entry for the small-scale mode and $C_1$ is an integration constant. The large-scale density perturbation in the radiation dominated era is outside the horizon and varies as $\delta_c \propto a^2$ at $a_{i}$. Plugging this into  Eq.~(\ref{eq:epsilon}), we obtain
\be
\epsilon = C_{1}\frac{1}{3}\delta_c(a, k_L)+C_{2}+C_{3}\log(a)\,,
\ee
where $C_2$ and $C_3$ are integration constants. We only keep the term $\propto \delta_{c} \propto a^{2}$ as by definition, $\epsilon$ vanishes as $\delta_c \rightarrow 0$. Thus, up to an overall normalization factor, we obtain the initial conditions,
\be
D_{i} = \log\left(\frac{a_{i}}{a_{H}}\right),\ \ \ \epsilon_i = \frac{1}{3}\delta_c(a_i, k_L) \,.
\ee

Using the above initial conditions, we can solve for $\epsilon$ and $D$ to obtain the linear growth response,
\be
\label{eq:dlnDWddelta}
R_D(k_L,a) \equiv \frac{\Delta\log D_{W}}{\Delta\delta_{c}} = \frac{\epsilon }{D \delta_c}\,,
\ee
where $R_D$ , defined above, depends on redshift and on the $k_L$ of the long wavelength mode $\delta_c$. 
\begin{figure}
    \centering
    \includegraphics[width=\columnwidth]{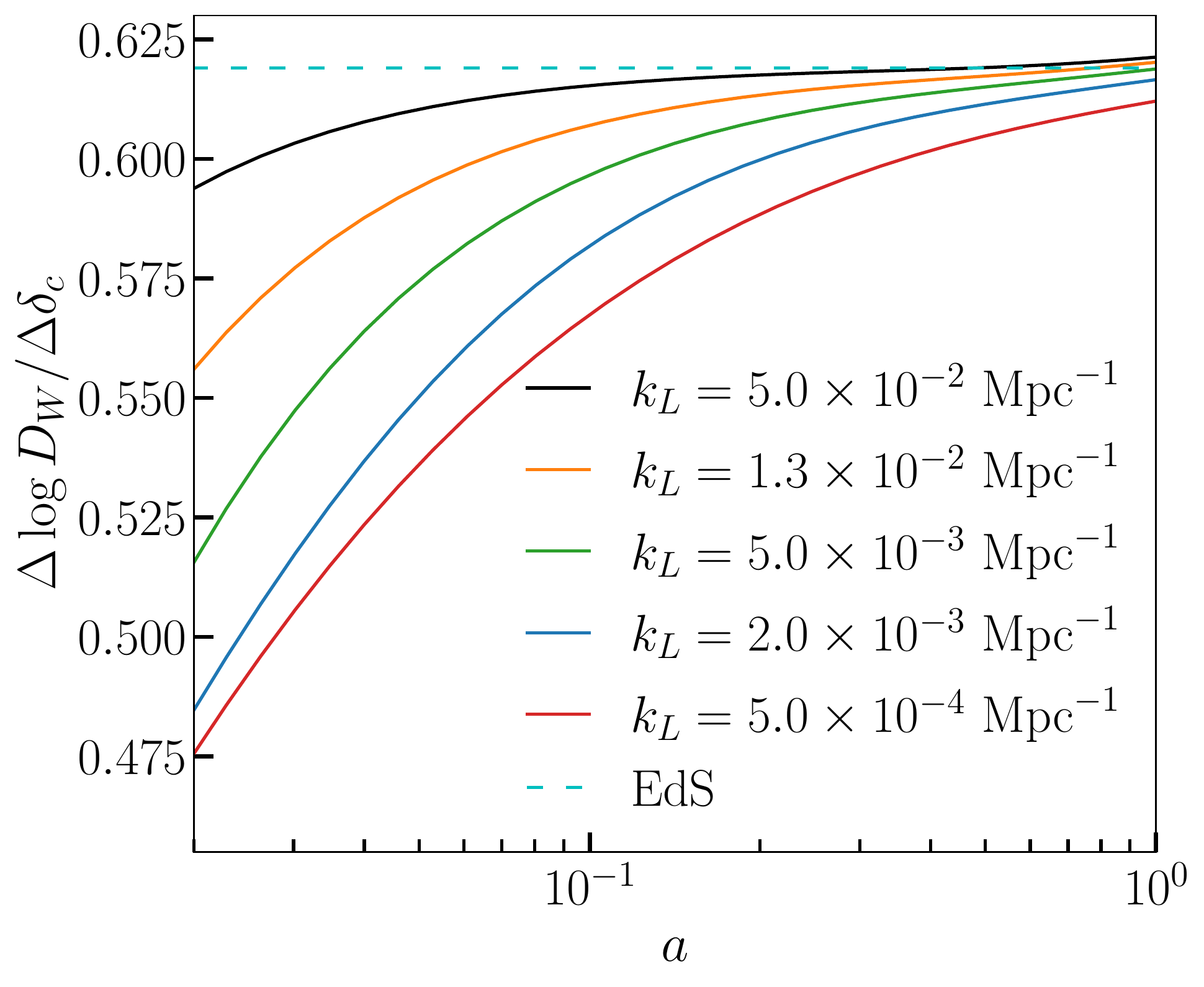}
    \caption{The response of the linear growth factor (Eq.~(\ref{eq:dlnDWddelta}) to the long wavelength modes shown in Fig. \ref{fig:deltac}. The dashed line is the same quantity computed in an exactly matter-dominated Separate Universe (= $13/21$).}
    \label{fig:growthresponse}
\end{figure}

Figure \ref{fig:growthresponse} shows the linear growth factor response as a function of the scale factor for Separate Universes built from different long-wavelength modes (i.e. different values of $k_L$) with $\Omega_{\mathrm{rad}}=3.7\times 10^{-4}$. It is apparent from Fig. \ref{fig:growthresponse} that even in the matter-dominated era this response differs from its counterpart in the exactly matter dominated (EdS) universe which is equal to $13/21$ (denoted by the dashed line in Fig. \ref{fig:growthresponse}).  This is due to the high value of the radiation density we have assumed here. For the standard amount of radiation composed of CMB photons at temperature $T_{cmb} = 2.725\ K$ and three massless neutrinos, the linear growth factor response at $z=0$ agrees with the EdS prediction to within $\lesssim 0.01\%$. 

\section{Analytical Results}
 \label{sec:analyticalresults}

 A broad goal of this work is to understand how free-streaming components such as neutrinos or photons affect the coupling between small-scale structure formation and large-scale density perturbations. At lowest order, this nonlinear coupling can be expressed in terms of the Separate Universe linear growth factor response, computed in the preceding section (Eq.~(\ref{eq:dlnDWddelta})). 
 
 The presence of free-streaming components introduces a scale dependence to the evolution of large-scale density perturbations $\delta_{c}(k_L,a)$, and thus causes the linear growth factor response to be dependent on $k_L$, unlike its counterpart in purely $\text{CDM}$ and $\Lambda\mathrm{CDM}$ cosmologies. From Fig.~\ref{fig:growthresponse}, we see that the linear growth factor response increases with increasing $k_L$. 
It is therefore instructive to look at the \textit{relative} linear growth factor response with respect to its asymptotic value at low $k_L$,   
\be
 \label{eq:relgrowth}
 R_{\textrm{rel}}(k_L, a) \equiv \frac{R_D(k_L,a)}{R_D(k_{L0}, a)} \, , 
 \ee
where $k_{L0} = 1 \times 10^{-4}\ \text{Mpc}^{-1}$ is a reference large-scale wave number chosen such that the linear growth factor response $R_{D}(k_L,a)$ is sufficiently close to its $k_L\rightarrow 0$ asymptotic value. We define the \textit{step} in the linear growth factor response as the asymptotic value of $R_{\textrm{rel}}$ in the limit of large $k_L$, 
\be
 \label{eq:stepdef}
 R_{\textrm{step}}(a) \equiv R_{\textrm{rel}}(k_{L}\gtrsim 0.06\ \text{Mpc}^{-1})\, .
\ee
In the following subsections we will discuss how the step in the linear growth factor response evolves with redshift and depends on the masses and energy densities of the free-streaming particles. We will also demonstrate that for massless particles, the $k_L$ dependence of the relative linear growth factor response is insensitive to the composition of the free-streaming component.
    
 \subsection{Dependence on particle mass and energy density}
 \label{ssec:Analytical_results_A}

  \begin{figure}
     \centering
     \includegraphics[width=\columnwidth]{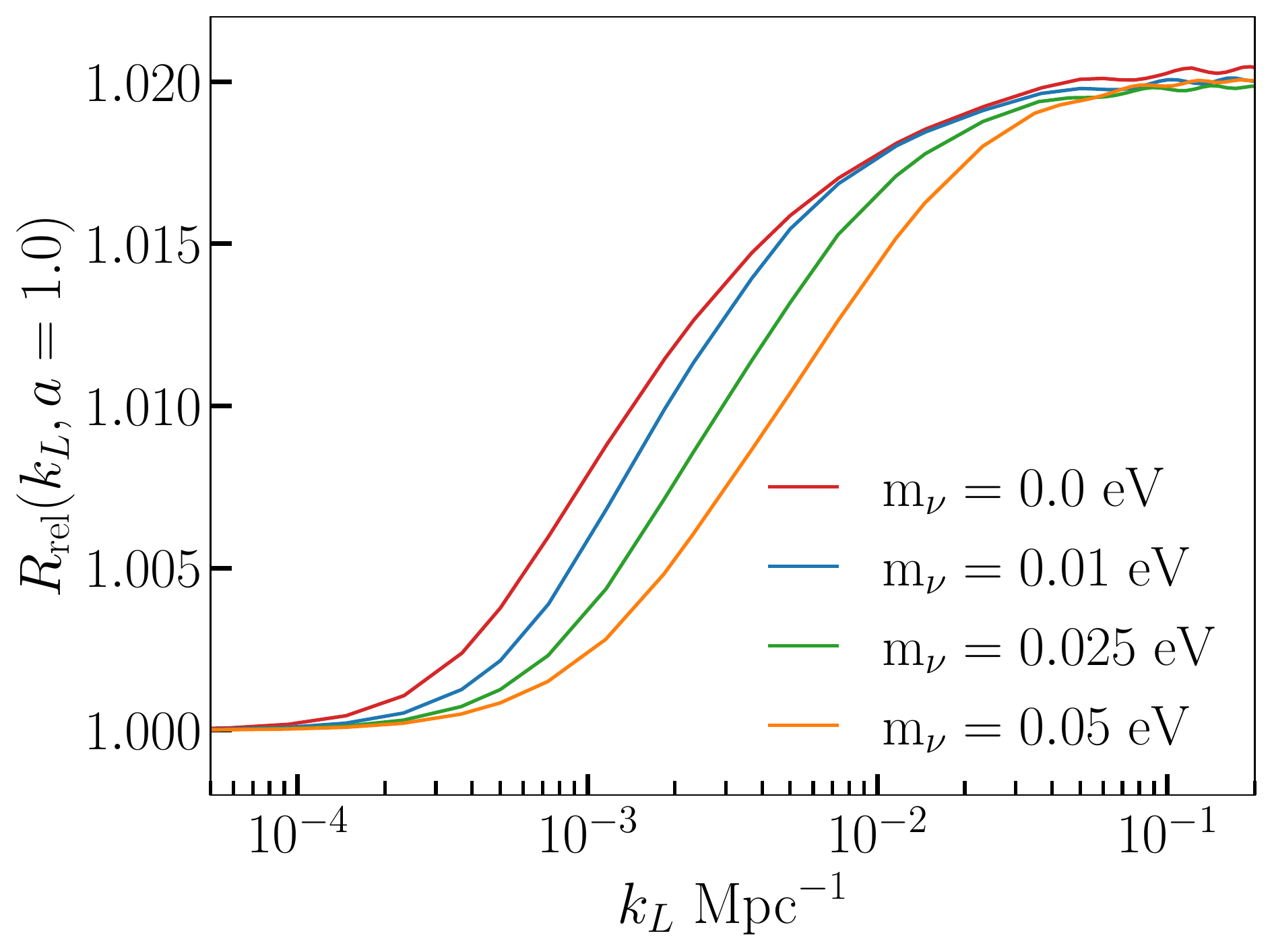}
     \caption{Relative linear growth factor response, Eq.~(\ref{eq:relgrowth}), for neutrinos of different masses but common temperature. The location of the step feature depends on the particle mass, in this case implemented as neutrinos with the standard temperature, $T_\nu \approx 1.7 \times 10^{-4}$~eV, and degenerate masses $0.05,\ 0.025,\ 0.01$ and $0.0~\text{eV}$. These curves do not have a common number of species nor energy density of neutrinos, the number of states was adjusted to produce the same asymptotic value at high $k_L$. }
     \label{fig:growthresponsemassdep}
 \end{figure}

 \begin{figure}
     \centering
     \includegraphics[width=\columnwidth]{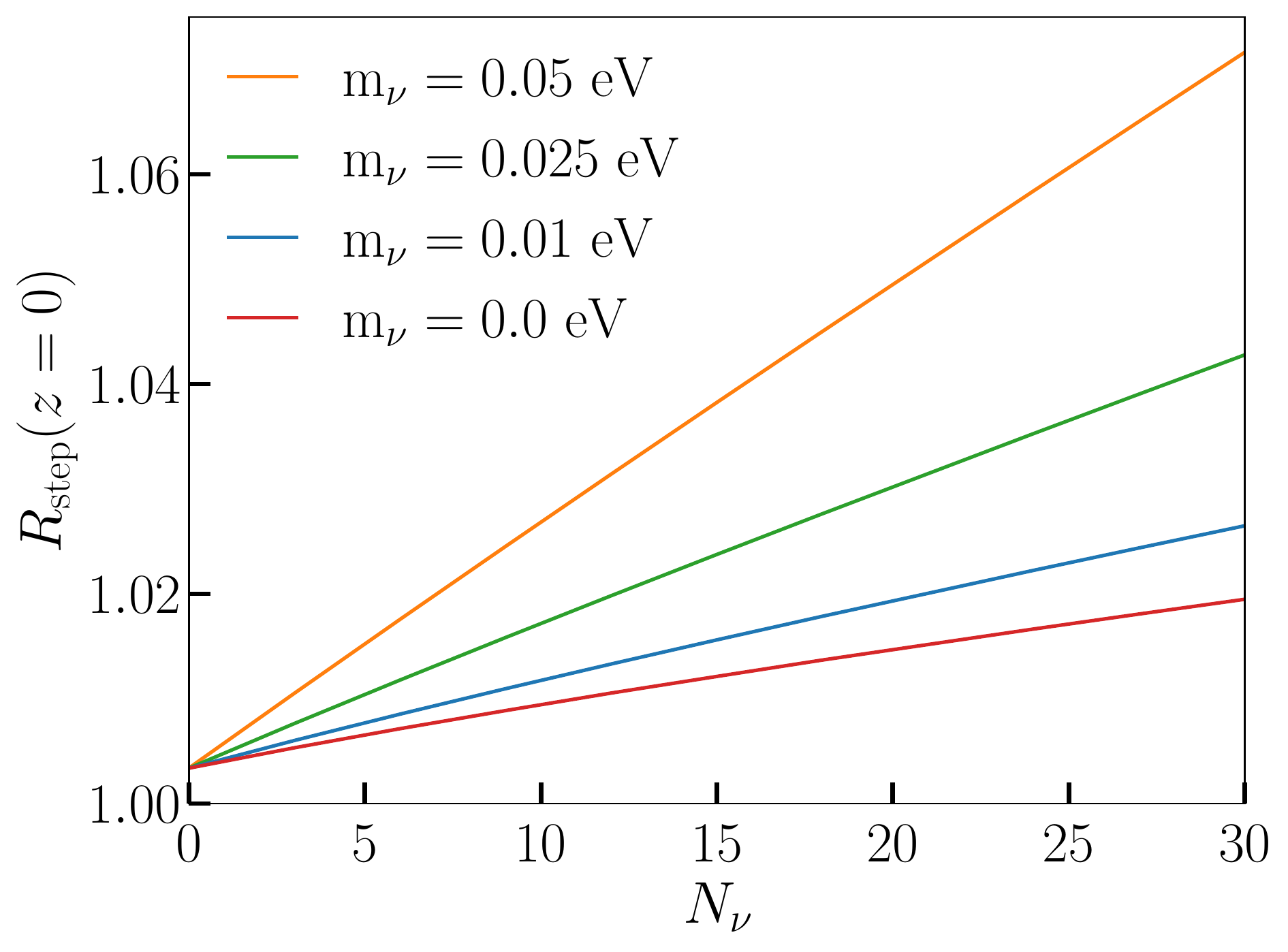}
     \caption{The amplitude of the step in the linear growth factor response, Eq.~\eqref{eq:stepdef} vs number of neutrino species $N_{\nu}$ at $z=0$ for neutrinos of degenerate mass $m_{\nu} = 0.05,\ 0.025,\ 0.01$ and $0.0$~eV.} 
     \label{fig4}
 \end{figure}
 
 \begin{figure}
     \centering
     \includegraphics[width=\columnwidth]{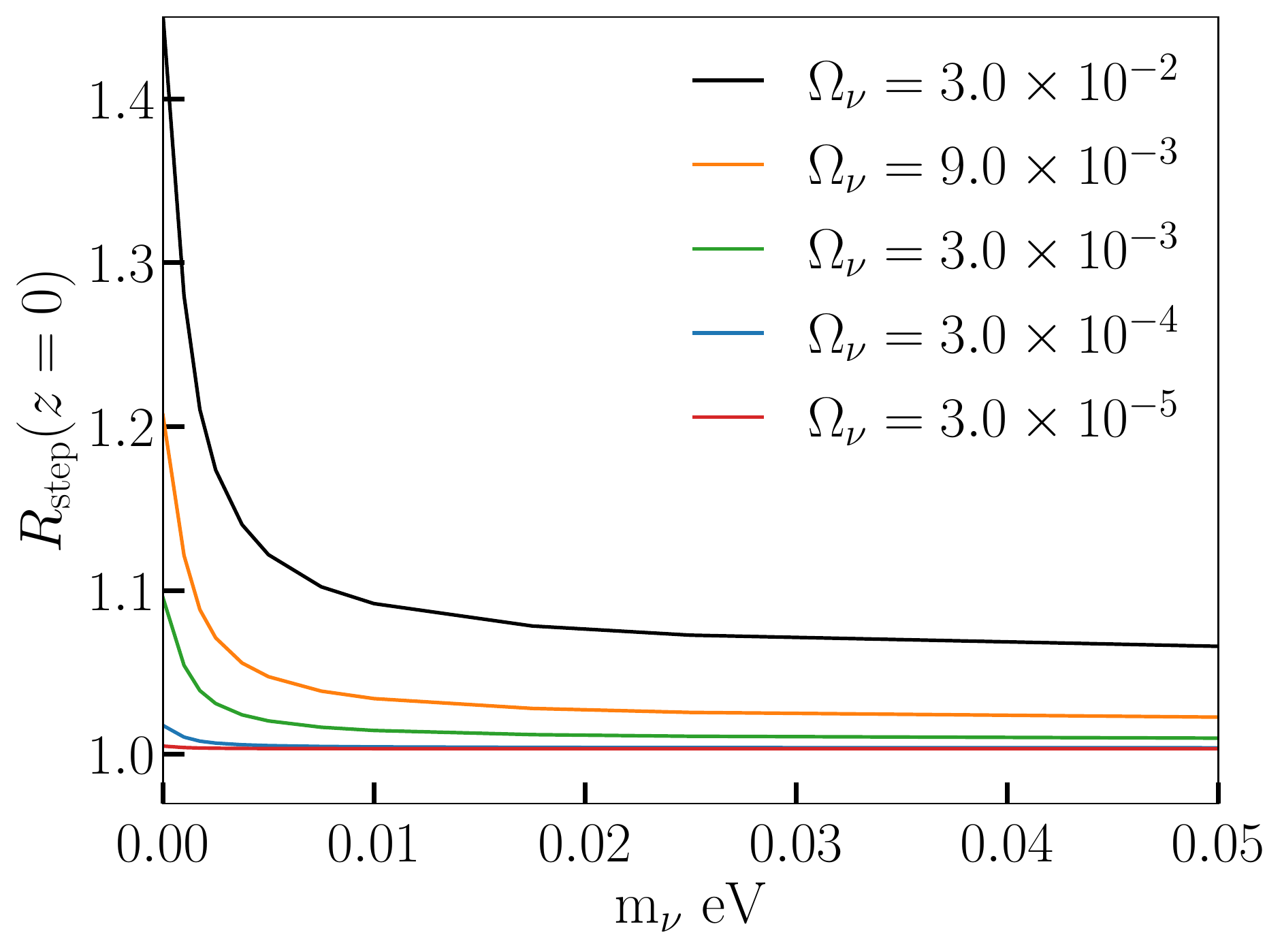}
     \caption{The amplitude of the step in the linear growth factor response, Eq.~\eqref{eq:stepdef} as a function of neutrino masses for fixed neutrino energy densities at $z=0$.} 
     \label{fig:growthresponsemassdep2}
\end{figure}

The relative linear growth factor response transitions from a lower value at small $k_L$ to a larger value at high $k_L$. The value of $k_L$ at which the relative linear growth factor response in Eq.~(\ref{eq:relgrowth}) starts increasing above $1$, depends on the free-streaming scale of the particles.  For particles with a typical speed $v_{th}$, the free-streaming scale in proper coordinates is given by \cite{lesgourgues2006massive},
 \be
 \lambda_{fs} \propto v_{th}/H\,.
 \ee
For massless neutrinos the free-streaming scale is just the Hubble scale, $c/H$, whereas for massive neutrinos $v_{th} \propto T_{\nu} / m_{\nu}$. Thus, at fixed temperature, the free-streaming scale is larger for neutrinos with smaller mass. 

Figure \ref{fig:growthresponsemassdep} shows the relative linear growth factor response  for neutrinos with masses $0.05,\ 0.025,\ 0.01$ and $0.0$~eV, computed at $z=0$ with a fixed neutrino background temperature. This plot shows that the relative growth response steps up at smaller scales for neutrinos of higher mass due to the decreasing free-streaming scale. For the plot in Fig. \ref{fig:growthresponsemassdep}, we have adjusted the neutrino abundances to match the asymptotic values of the growth response at high $k_L$ in order to clearly illustrate the changes in the location of the growth response step.

 The step in the linear growth factor response increases with an increase in the energy density of light particles (with $\Omega_{c}$ constant). In Fig. \ref{fig4} we show the amplitude of the linear growth response step for both massive and massless neutrinos for different values of $N_{\nu}$ and $m_{\nu}$ at redshift $z=0$. The size of the response step at a given redshift increases with increasing energy density ($\sim m_\nu N_\nu$ for $m_\nu \gtrsim 0.01$~eV) of the light particles. Figure \ref{fig4} shows a non-trivial linear growth response step for $N_{\nu} = 0$. We interpret this as the growth factor response step due to the presence of photons. Photons, like neutrinos, are free streaming at late times (after recombination) and would be expected to have a similar effect on the growth factor response as massive neutrinos. We will verify this in Sec. \ref{ssec:composition}.

For a fixed energy density of free-streaming particles at a given redshift, the asymptotic value of the growth response step at high $k_L$ also depends on the mass of the free-streaming particles. This is evident from Fig. \ref{fig:growthresponsemassdep2} where we have plotted the amplitude of the growth response step as a function of the neutrino mass for fixed neutrino energy density at $z=0$. Notably, fixing the energy density in neutrinos of different masses at $z=0$ does not fix the energy densities at higher redshifts, where the neutrino momentum can contribute significantly to the neutrino energy. Fig. \ref{fig:growthresponsemassdep2} shows that neutrinos of low mass with the same energy density at $z=0$ generate a bigger growth response than their higher mass counterparts. We attribute this to the fact that, with fixed $\Omega_{\nu}$ at $z=0$, low mass neutrinos have a higher energy density at earlier times than neutrinos of higher masses.

\subsection{Dependence on composition}
\label{ssec:composition}
 \begin{figure}
     \centering
     \includegraphics[width=\columnwidth]{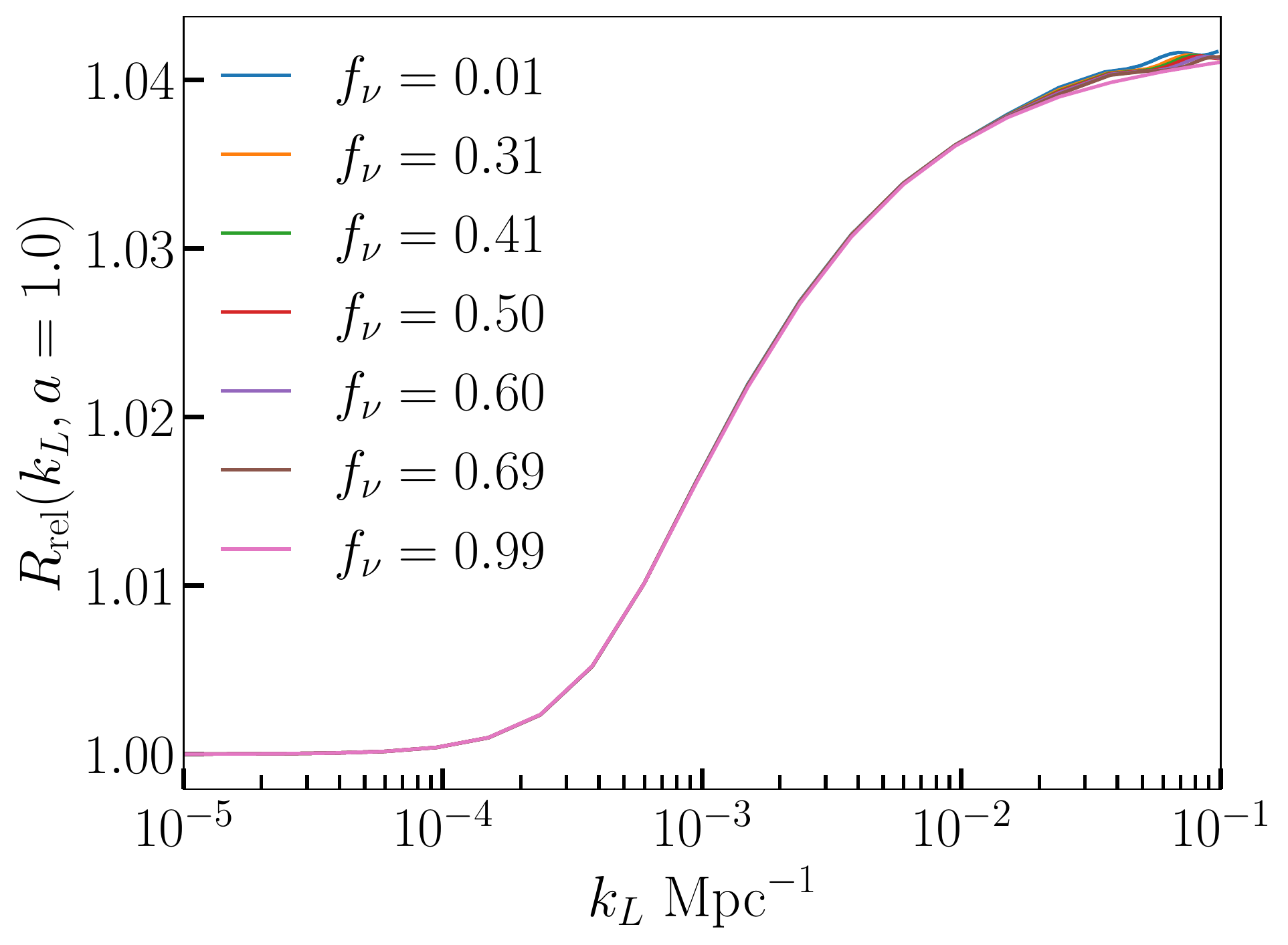}
     \caption{Relative linear growth factor response, Eq.~(\ref{eq:relgrowth}), for different massless neutrino densities with $\Omega_{rad,0} = 1.01\times 10^{-3}$. Here $f_{\nu}$ is the fraction of the radiation energy density carried by massless neutrinos, $f_\nu = \Omega_\nu/(\Omega_\nu+\Omega_\gamma)$. The fractional difference between the different curves is $\lesssim 5\times10^{-4}$. }
     \label{fig6}
\end{figure}

Since we are studying the gravitational effects of large-scale perturbations in radiation, we expect that the particular composition of the radiation (e.g. if it is photons, massless neutrinos, or some other relativistic particle) is unimportant. Yet, in our Universe photons and neutrinos have different cosmological histories owing to different decoupling times. Specifically, since we start our solutions to $D_W$ and $\epsilon$ deep in the radiation dominated era when photon and neutrino perturbations evolve differently, one may worry that this could cause $D_W$ to depend on composition. We study this by considering the growth response for different $k_L$ in the presence of different photon and massless neutrino energy densities producing the same total radiation energy density. Figure \ref{fig6} shows the $k_L$ dependence of the relative growth response at $z=0$, for difference fractions of $\Omega_{\nu}$ and $\Omega_\gamma$ at fixed $\Omega_{rad} = 1.01\times 10^{-3}$. From this plot, it follows that there is no significant difference between the effect of photons and that of massless neutrinos on the response of small-scale structure formation to large-scale CDM density perturbations.

\subsection{Dependence on redshift}

\begin{figure}
     \centering
     \includegraphics[width=\columnwidth]{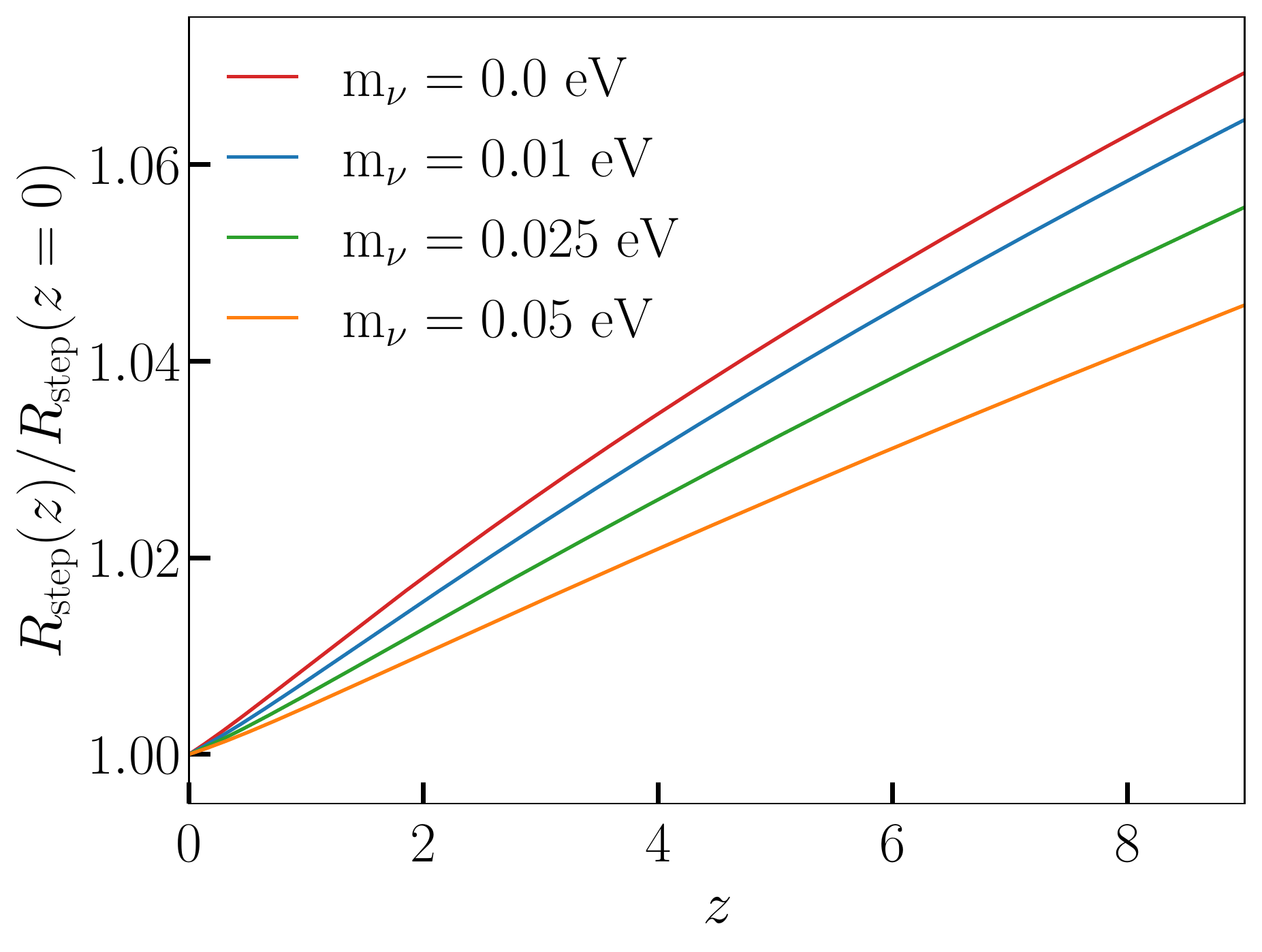}
     \caption{The step in the linear growth factor response normalized to unity at $z = 0$ plotted as a function of redshift for 28 species of neutrinos with different masses. }
     \label{Massive_massless_comparison}
\end{figure}

Figure \ref{Massive_massless_comparison} shows the amplitude of the step in the linear growth factor response between large and small scales evaluated using Eq. \eqref{eq:relgrowth} normalized to unity at redshift $z = 0$. It is evident from Fig. \ref{Massive_massless_comparison} that the amplitude of the step depends more strongly on the redshift in a universe with massless neutrinos than in a universe with massive neutrinos. This can be attributed to the fact that at late times the energy density of massless neutrinos falls faster with the expansion of the universe than that of massive neutrinos. 

\section{Separate Universe simulations}
\label{sec:Nbody}

 We performed N-body simulations in the Separate Universe to obtain the response of structure formation at small scales to the long-wavelength density perturbations in the non-linear regime. We used a modified version of the N-body code Gadget-2 \cite{springel2005cosmological}, which reads in a list of tabulated values for $(a_W, H_W)$, and interpolates from this list rather than computing the expansion history using cosmological parameters. Our simulations began at a Separate Universe scale factor $a_{Wi}$ corresponding to a redshift of $z=49$ in the global universe. We constructed initial conditions with an initial power spectrum 
 \be
 P_W(k,a_{i}) = \left[\frac{D_{Wi}}{D_{0}}\right]^2 P(k, a_{0})
 \ee
 where $a_0 = 1$ is the scale factor today, $D_{Wi}$ is the Separate Universe linear growth factor at the initial redshift, and $D_{0}$ is the linear growth factor in the global universe at $a_{0} = 1$. $P(k,a_{0})$ is the linear CDM + baryon power spectrum at $a_{0}$, computed using CLASS with the cosmological parameters given in Table \ref{cosmology}, including 28 massless neutrinos. $P_{W}(k,a_{i})$ is initial power spectrum in the Separate Universe. We then generated realizations of Gaussian random fields, which we evolved to the initial scale factor $a_{Wi} = 0.02$ using second order Lagrangian Perturbation Theory (2LPT) \cite{Crocce:2006ve} in the matter-radiation dominated Separate Universe \cite{chiang2018scale}. We carried out the simulations until the time corresponding to $a_{W} = a_{0}(1-\delta_{c0}/3)$ where $a_0 = 1$, i.e. up to the same \textit{physical} present time as the global universe. 
 
 We identified halos in our simulations at redshifts $z=0$ and $z=1$ using the Rockstar halo finder \cite{behroozi2012rockstar}. Rockstar uses a friends-of-friends algorithm to find halos, and then assigns halo masses using a spherical overdensity calculation around each halo's center of mass. The friends-of-friends halos are grown radially outward until the spherically averaged density crosses a threshold, defined by $\rho_{\mathrm{TH}} = \bar{\rho}_{c} \Delta$. We used the virial threshold, so that $\Delta$ is redshift dependent and defined such that $\rho_{\mathrm{TH}}$ is the virial density at a given redshift. While Rockstar computes the spherical overdensity in Separate Universe comoving coordinates, the density thresholds are computed with respect to the global cosmology, so the threshold factor $\Delta$ must be adjusted for the Separate Universe halos. If $\Delta$ and $\Delta_W$ set the  density thresholds in the global universe and Separate Universe respectively,
 \be
 \overline{\rho}\Delta = \overline{\rho}_{W}\Delta_{W} \implies \Delta_{W} = \frac{\Delta}{1+\delta_{c}(a)} \approx \Delta ( 1-\delta_{c}(a))
 \ee
We also used a modified version of the spherical overdensity mass assignment, which computes continuous values for the halo masses rather than the discrete masses from counting particles (see \cite{li2016separate, Jamieson:2018biz} for details).
 
We ran our simulations in a box of comoving length $700\ \text{Mpc}$ with $640^3$ particles.  We chose to restrict our halo catalogs to halos containing more than $100$ particles and quote our results for halos containing more than $374$ particles, corresponding to a halo mass of $M = 2 \times 10^{13}~\mathrm{M}_\odot$.
 
 We ran pairs of overdense and underdense Separate Universe simulations with $\delta_{c0} = \pm 0.01$ for two long-wavelength modes spanning the step visible in Fig. \ref{fig:growthresponsemassdep}, $k_L = 5 \times 10^{-2}\ \text{Mpc}^{-1}$ and $k_L = 5 \times 10^{-4}\ \text{Mpc}^{-1}$, we will refer to these wave numbers as $k_\downarrow$ and $k_\uparrow$, respectively. We carried out $110$ such pairs of simulations and bootstrap averaged over any quantities measured from them to estimate statistical errors. For each pair of overdense and underdense simulations, we used the same initial realization of the Gaussian random field so that cosmic variance largely canceled in the Separate Universe responses we measured \cite{chiang2018scale}.
 
 Our simulation box length of $700\ \text{Mpc}$ is within the Hubble scale at all times (the comoving Hubble scale is $\approx 1.12$~Gpc at $a_{Wi}$ to $\approx 4$~Gpc at $z=0$) so that we could safely neglect the clustering of radiation within our simulations, since it only affects super-Hubble scales.

\section{Results from N-body simulations}
\label{sec:results from simulations}

\subsection{Power Spectrum Response}

\begin{figure}
    \centering
    \includegraphics[width=\columnwidth]{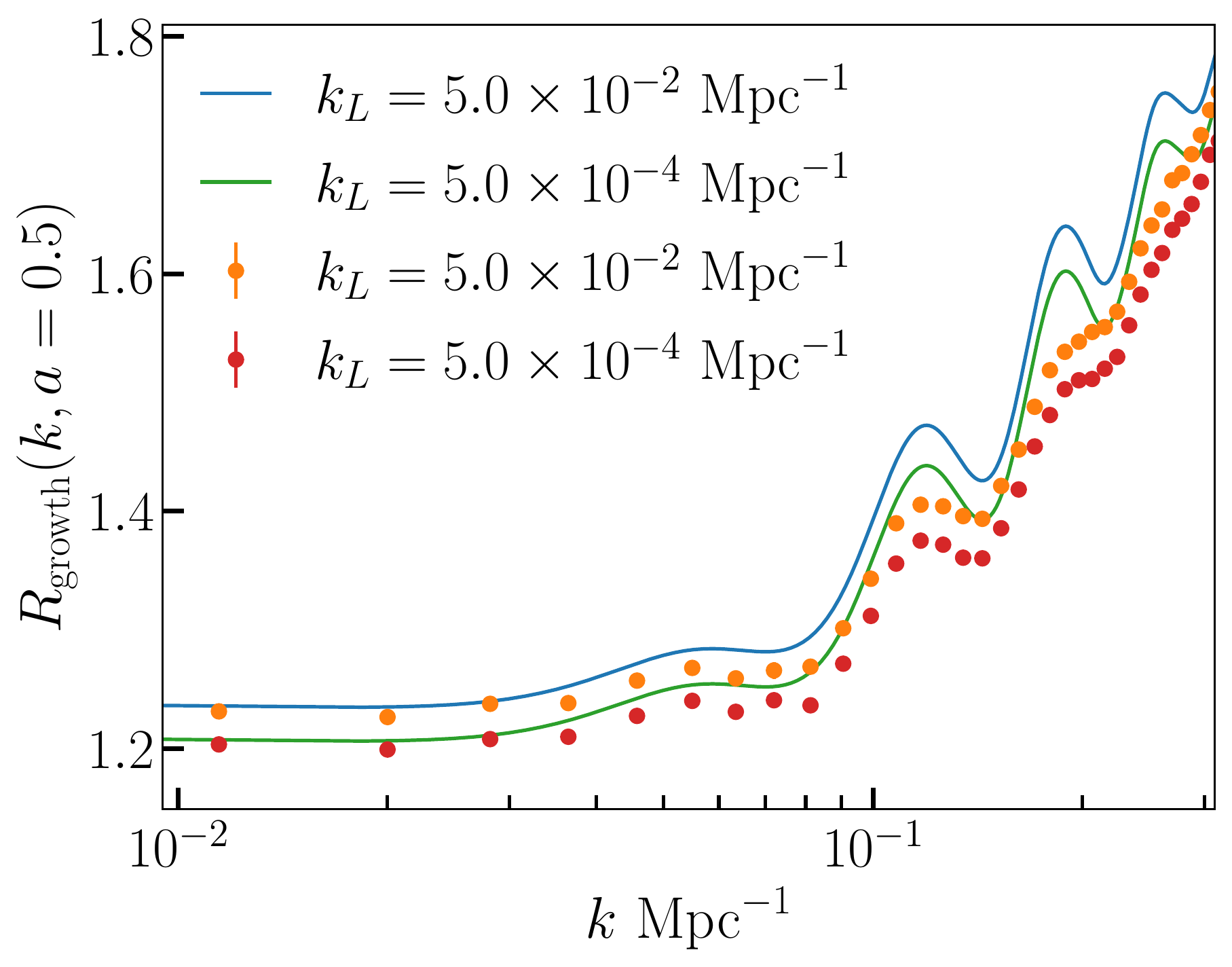}
    \caption{The growth contribution to the power spectrum response, $R_{\textrm{growth}}$, for different long-wavelength modes at $z=1$. The points show results from $N$-body simulations in the Separate Universe, while the curves show predictions from $1$-loop perturbation theory. }
    \label{fig9}
\end{figure}

\begin{figure}
    \centering
    \includegraphics[width=\columnwidth]{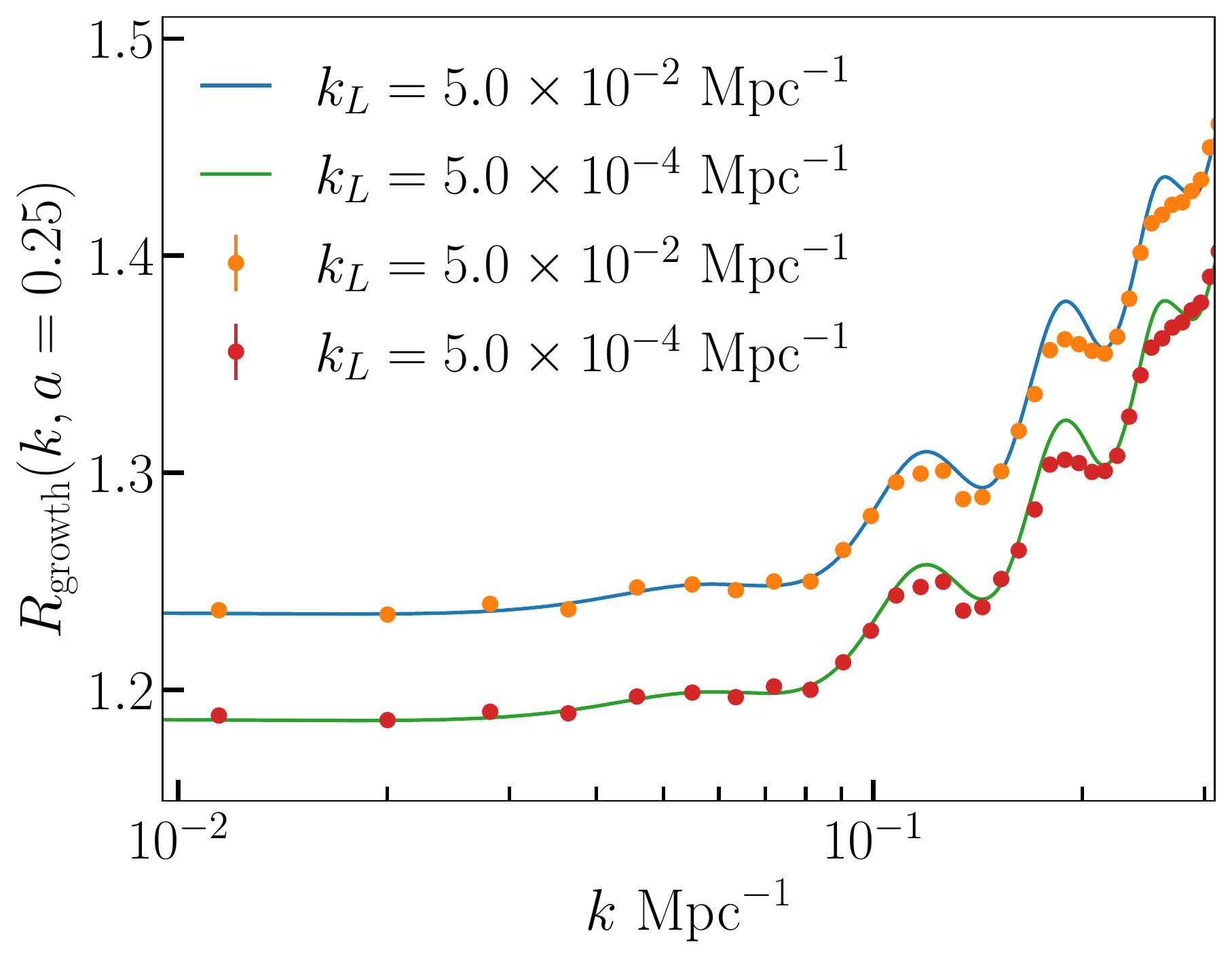}
    \caption{The growth contribution to the power spectrum response, $R_{\textrm{growth}}$,  for different long-wavelength modes at $z=3$. The points show results from $N$-body simulations in the Separate Universe, while the curves show predictions from $1$-loop perturbation theory. }
    \label{fig10}
\end{figure}

The small-scale matter power spectrum in presence of a background large-scale CDM density perturbation differs from the global matter power spectrum. We define the power spectrum response $R_{\textrm{tot}}$ by the equation,
\begin{eqnarray}
\frac{\Delta P}{P} = R_{\textrm{tot}}(k, k_L)\delta_{c}(k_L)\,,
\end{eqnarray}
where $\Delta P(k, k_L)$ is the difference between power spectra in the separate and global universes, computed at fixed proper wave number $k/a$. At leading order, $R_{\textrm{tot}}$ is independent of $\delta_{c}$ and can be computed using the overdense and underdense pairs of Separate Universe simulations. 

The power spectrum response $R_{\textrm{tot}}$ can also be obtained from the squeezed limit of the CDM + baryon bispectrum spectrum averaged over the angle between the short and long mode wave vectors \cite{Chiang:2014oga},
\begin{eqnarray}
\lim_{k_{L}\to\infty} B(k,k',k_{L}) \approx R_{\textrm{tot}}P(k_{L})P(k)\,.
\end{eqnarray}
The power spectrum response $R_{\textrm{tot}}$ can be decomposed as, 
\be
R_{\textrm{tot}}(k, k_L) = R_{\textrm{growth}}(k, k_L) + R_{\textrm{dilation}}(k) + R_{\overline{\rho}}
\ee
where $R_{\textrm{growth}}$ denotes the fractional change in the matter power spectrum at a fixed comoving $k$; $R_{\textrm{dilation}}$ is the contribution due to the difference in comoving $k$ in the global and Separate Universes; and $R_{\overline{\rho}}$ is the contribution due to the difference in the background matter densities against which density contrasts are computed in both the cosmologies \cite{Li:2014sga}. The terms $R_{\textrm{dilation}}$ and $R_{\textrm{growth}}$ are artifacts of the Separate Universe construction and can be determined from the global universe power spectrum,
\be
R_{\textrm{dilation}}(k) = -\frac{1}{3} \frac{d\log\left( k^3 P \right)}{d\log k}\,,
\ee
\be
R_{\overline{\rho}} = 2\,.
\ee
The dynamical quantity $R_{\textrm{growth}}$ encodes the different growth rates of matter perturbations in presence of a long-wavelength mode. We can compute $R_{\textrm{growth}}$ using pairs of Separate Universe simulations by computing the power spectra in the overdense and underdense universes for the same comoving $k$. We divided our simulation box into a grid with $640^3$ cells and obtained the density fluctuations in position space using a cloud-in-cell density assignment. We then Fourier transformed the density fluctuation field using FFTW3 \cite{frigo2005design}. $R_{\textrm{growth}}$ is then given by 
\begin{eqnarray}
\hat{R}_{\text{growth}} = \frac{\hat{P}(k,a|\delta_{c0}^{+}) - \hat{P}(k,a|\delta_{c0}^{-})}{2\delta_{c}(a)\hat{P}(k,a|\delta_{c0}=0)}\,,
\end{eqnarray}
where $\delta_{c0}^{\pm} = \pm 0.01$ are the values of $\delta_c$ at redshift $z=0$ in the overdense and underdense simulations, and $\hat{P}$ denotes the matter power spectrum computed from the simulations. We approximated $P(k,a|\delta_{c}=0)$ as the average power spectra from the overdense and underdense simulations, which differs from the true global power spectrum only at order $\mathcal{O}(\delta_c^2)$. 

Figures \ref{fig9} and \ref{fig10} show the power spectrum growth response $R_{\textrm{growth}}$ computed at $z=1,3$ and bootstrap averaged over $110$ pairs of Separate Universe simulations. The solid curves in those plots show the power spectrum response computed using 1-loop perturbation theory. To compute the  1-loop power spectrum response we have,
\begin{eqnarray}
\frac{d\log P_{W,1-loop}}{d\delta_{c}} = \frac{d\log P_{1-loop}}{d\log D}\left[\frac{d\log D_{W}}{d\delta_{c}}\right]\,,
\end{eqnarray}
with $d\log D_{W}/d\delta_{c}$ being the linear growth factor response $R_{D}(k_L,a)$ defined in Eq.\eqref{eq:dlnDWddelta}. $ P_{1-loop} = P_{lin}+2P_{13} +P_{22} $ where $P_{lin} \propto D^2$ is the linear power spectrum and $P_{22}, P_{13} \propto D^4$ are the non-linear corrections at 1-loop. Therefore, 

\begin{eqnarray}
\frac{d\log P_{1-loop}(k,a)}{d\log D(a)} = 2\left[1+\frac{P_{22}(k,a)+2P_{13}(k,a)}{P_{lin}(k,a)}\right]\,.
\end{eqnarray}

Figures \ref{fig9} and \ref{fig10} show with high statistical significance that $R_{\textrm{growth}}$ is dependent on $k_{L}$. This dependence on $k_L$ arises due to the scale-dependent evolution of $\delta_{c}$ in the presence of radiation. We also see that our 1-loop calculation of the power spectrum agrees with the power spectrum response computed using Separate Universe simulations when the small-scale wave number approaches large scales, $k \lesssim 0.3\ \text{Mpc}^{-1}$. This is exactly the regime in which 1-loop perturbation theory has predictive power. Also, the calculation from 1-loop perturbation theory agrees better at $z=3$ with the results from Separate Universe simulations than at $z=1$, which is also expected since the evolution of matter density fluctuations becomes more nonlinear with time.

\subsection{Scale-dependent halo bias}

\begin{figure}
    \centering
    \includegraphics[width=\columnwidth]{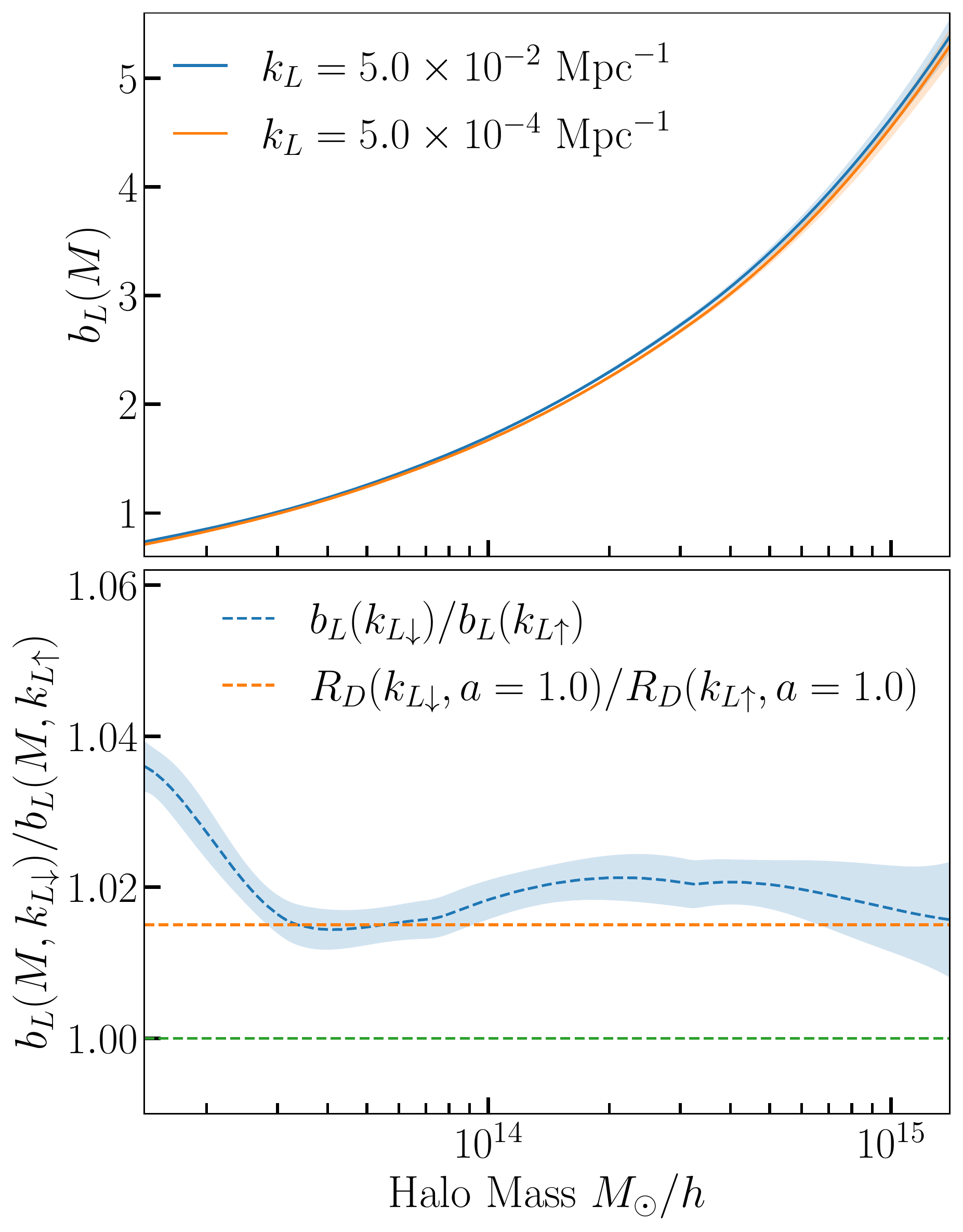}
    \caption{(Top) Cumulative Lagrangian bias as a function of halo mass at $z=0$. (Bottom) Relative Lagrangian bias between $k_{L\downarrow}$ and $k_{L\uparrow}$ as a function of halo mass at $z=0$. The dashed orange line shows the ratio of the linear growth factor responses $R_{D}(k_{L\downarrow})/R_{D}(k_{L\uparrow})$.}
    \label{fig12}
\end{figure}

\begin{figure}
    \centering
    \includegraphics[width=\columnwidth]{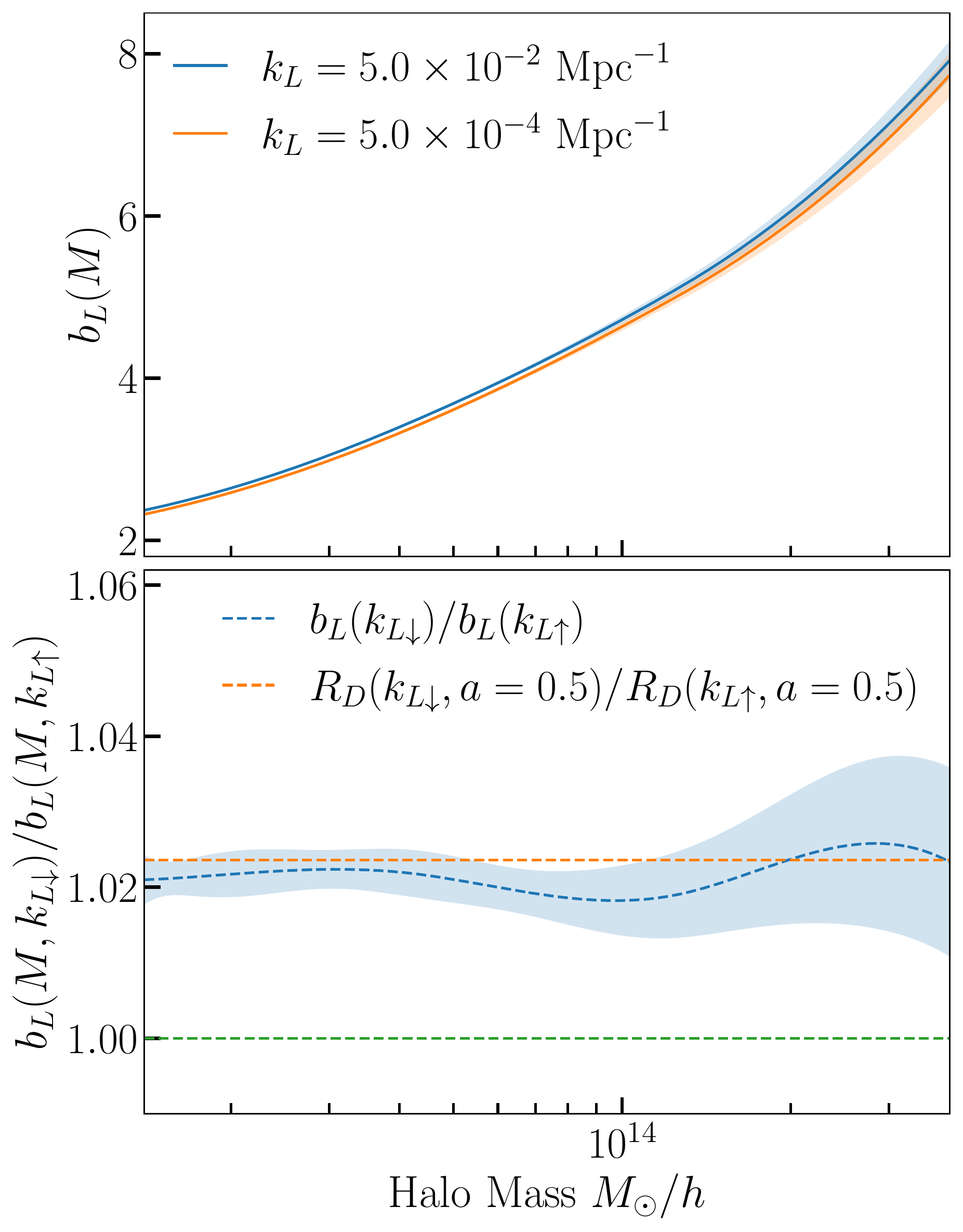}
    \caption{(Top) Cumulative Lagrangian bias as a function of halo mass at $z=1$. (Bottom) Relative Lagrangian bias between $k_{L\downarrow}$ and $k_{L\uparrow}$ as a function of halo mass at $z=1$. The dashed orange line shows the ratio of the linear growth factor responses $R_{D}(k_{L\downarrow})/R_{D}(k_{L\uparrow})$.}
    \label{fig17}
\end{figure}

We now turn to the dependence of the halo mass function in the Separate Universe on $\delta_{c}$. We define the Separate Universe \textit{Lagrangian response bias} as the variation of the cumulative halo mass function in the Separate Universe with respect to the long-wavelength density fluctuation $\delta_{c}$,
\begin{eqnarray}
b_{L}(M) = \frac{d\log n_W(\geq M)}{d\delta_{c}} \,,
\end{eqnarray}
where $n_W(\geq M)$ is the comoving number density of halos with mass larger than $M$ in the Separate Universe. Calculating the Lagrangian response bias gives a calibration of the standard clustering bias of halos \cite{li2016separate}. The response of $ n(\geq M)$ to $\delta_{c}$ arises primarily as a result of a small change in the mass of halos within a particular mass bin. Lower variance results are obtained using the abundance matching method given in \cite{li2016separate}. A straightforward way to understand the abundance matching method is to observe that, 
\begin{eqnarray}
\left(\frac{\Delta \log n_W}{\Delta\log M}\right)_{\delta_{c}}\left(\frac{\Delta \log M}{\Delta\delta_{c}}\right)_{n}\left(\frac{\Delta\delta_{c}}{\Delta\log n_W}\right)_{M} = -1\,,
\end{eqnarray}
which implies,
\begin{eqnarray}
b_{L}(M) = \frac{n_{\log M}(\log M)s(\log M)}{n(\log M)} \,,
\end{eqnarray}
where  
\begin{align}
s(\log M) = \left(\frac{\Delta \log M}{\Delta\delta_{c}}\right)_{n} \, ,\\ 
n_{\log M}(\log M) = -\frac{dn(\geq M)}{d\log M} \, .
\end{align}
To obtain these quantities, we bootstrap resampled $110$ pairs of Separate Universe simulations and combined the halo catalogs (sorted in halo mass) of the overdense and underdense simulations respectively. We then estimated $s(M)$ as,
\begin{eqnarray*}
s_{i}(\log M_{i})=\frac{\log M_{i}^{+}-\log M_{i}^{-}}{2}\,,
\end{eqnarray*}
where $M_{i}^{\pm}$ are the masses of the $i^{th}$ most massive halos in the overdense and underdense halo catalogs respectively and $M_{i} = \sqrt{M_{i}^{+}M_{i}^{-}}$. We used a smoothing spline  to estimate the threshold mass shift $\hat{s}(\log M)$ and the cumulative (Lagrangian) halo mass function $\hat{n}(\log M)$. We also estimated the halo mass function $\hat{n}_{\log M}$ by taking the spline derivative of $\hat{n}(\log M)$. We then estimated the Lagrangian response bias as, 
\begin{eqnarray}
\hat{b}_{L}(M) = \frac{\hat{n}_{\log M}(\log M)\hat{s}(\log M)}{\hat{n}(\log M)}\,.
\end{eqnarray}
We repeated this procedure with $1000$ resamplings to obtain the mean and the error in the mean.

As remarked before, the growth of $\delta_{c}$ in the presence of massless neutrinos (or any energy component possessing a Jeans scale) becomes scale dependent. This makes the halo bias dependent on the large-scale $k_{L}$ \cite{Parfrey:2010uy, LoVerde:2014pxa}. Figure \ref{fig12} (top) and Figure \ref{fig17} (top) show the cumulative Lagrangian bias computed at redshifts $z=0$ and $z=1$, respectively, using abundance matching for $110$ pairs of overdense and underdense simulations at $k_{L} = 5\times 10^{-4}\ \text{Mpc}^{-1}$ and $k_L = 5 \times 10^{-2}\ \text{Mpc}^{-1}$. Figures \ref{fig12} (bottom) and \ref{fig17} (bottom) plot the ratio of the cumulative Lagrangian biases at the two values of $k_L$ as a function of the halo mass. Figures \ref{fig12} (bottom) and \ref{fig17} (bottom) demonstrate with high statistical significance that this ratio differs from unity, thus implying the dependence of $b_L(M)$ on $k_{L}$.

\section{Bias Models}
\label{sec:bias models}
It is instructive to compare the results of the halo bias obtained using N-body Separate Universe simulations to models of the halo bias developed in the literature. This way, one can critically evaluate the assumptions behind each of these models and possibly improve them, shedding more light on the non-linear process of halo formation in the presence of neutrinos. According to the simplest such model, the halo mass function can be regarded as a universal function of the matter power spectrum. In this case, the only dependence of the Lagrangian bias on $k_{L}$ arises from the growth response of the power spectrum \cite{chiang2018scale, Chiang:2018laa}\footnote{Note that for dark energy isocurvature and massive neutrino cosmologies this approach has been shown to give nearly identical predictions for the scale dependence of the halo bias as performing spherical collapse in the Separate Universe \cite{LoVerde:2014pxa, Chiang:2017vuk, Jamieson:2018biz}.}. In the Separate Universe picture, we would therefore have,
\begin{eqnarray}
b_{L}(M) = \frac{d\log n_W[M; P_{W}]}{d\delta_{c}} \propto \frac{d\log D_{W}}{d\delta_{c}}(k_{L})\,,
\end{eqnarray}
regardless of whether $P_{W}$ is the linear or non-linear power spectrum. Following this model, the scale dependence of the Lagrangian response bias at leading order should follow the scale dependence of the Separate Universe linear growth factor response,
\begin{eqnarray}
\frac{b_{L}(M,k_{L\downarrow})}{b_{L}(M,k_{L\uparrow})} = \frac{R_{D}(k_{L\downarrow})}{R_{D}(k_{L\uparrow})}\,.
\end{eqnarray}
Figures \ref{fig12} (bottom) and \ref{fig17} (bottom) show our N-body simulations results for $b_{L}(M,k_{L\downarrow})/b_{L}(M,k_{L\uparrow})$ at redshifts $z=0$ and $z=1$ respectively. We have compared these with the predictions from this model (shown by the orange dashed line in both figures). The results of our Separate Universe simulations agree, within high statistical significance, with the predictions of this model. This agreement is stronger at $z=1$ that at $z=0$.

\section{Effects on Observables}
\begin{figure}
    \centering
    \includegraphics[width=\columnwidth]{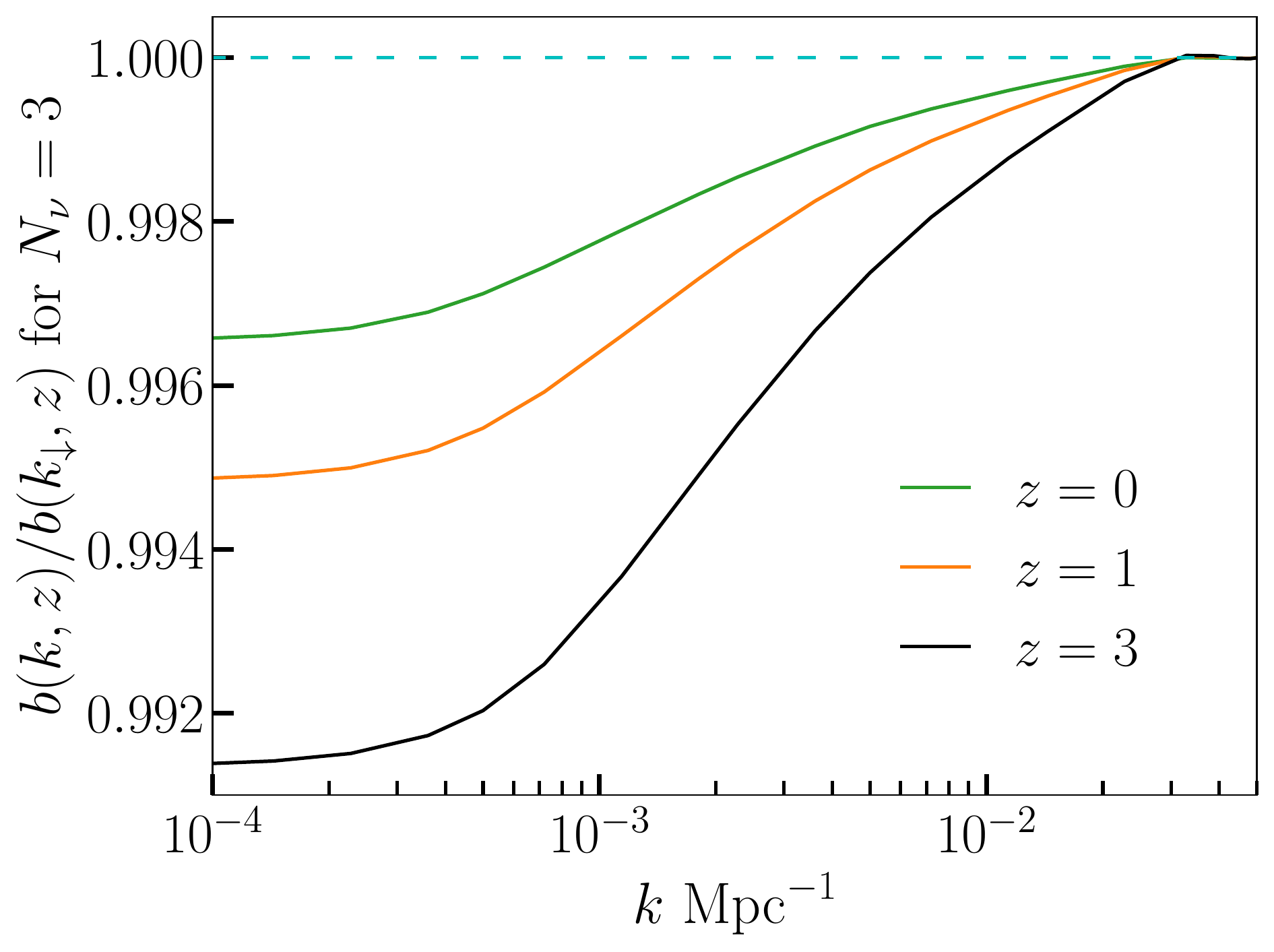}
    \caption{Plotted is the scale dependence of the Eulerian halo bias caused by large-scale perturbations in radiation composed of CMB photons and three massless neutrinos. This quantity is shown at several different redshifts for a fixed Eulerian bias $b = 2$ at $k_{\downarrow} = 5 \times 10^{-2}\ \mathrm{Mpc}^{-1}$.}
    \label{fig:Pkofk}
\end{figure}

\label{sec:effects on observables}
\begin{figure}
    \centering
    \includegraphics[width=\columnwidth]{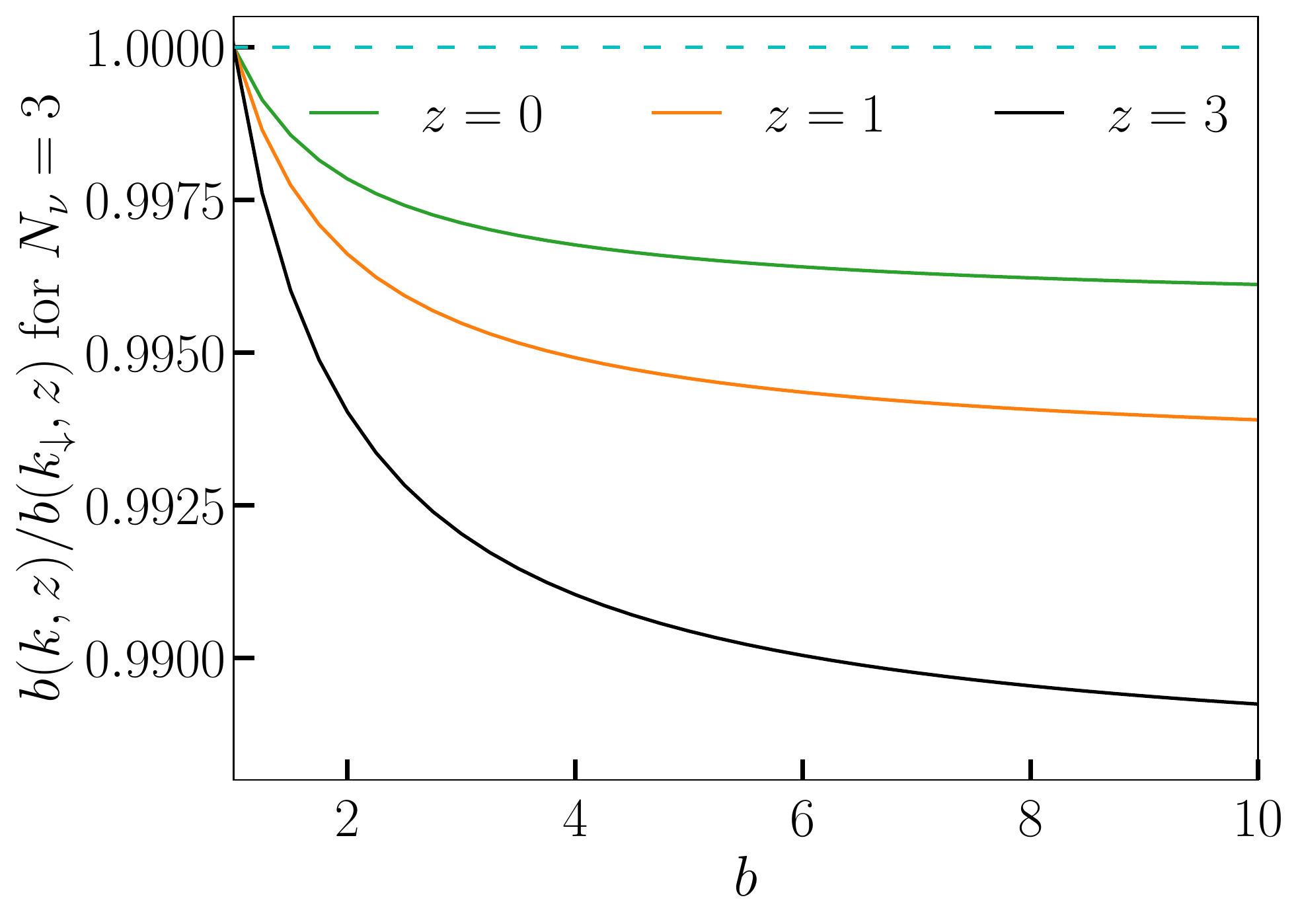}
    \caption{The overall amplitude of the change in the Eulerian halo bias between $k_\downarrow = 5 \times 10^{-2}\ \text{Mpc}^{-1}$ and $k_\uparrow = 5\times 10^{-4}\ \text{Mpc}^{-1}$, plotted in Fig. \ref{fig:Pkofk}, shown as a function of $b$, the Eulerian bias at $k_\downarrow$.}
    \label{fig:Pkofb}
\end{figure}

The scale-dependent effects due to free-streaming particles, such as those in the responses of power spectra and the halo mass function, will be present in the observables measured by cosmological surveys. For the halo auto power spectrum, the halo bias will acquire a scale-dependence due to the large-scale radiation perturbations, 
\begin{eqnarray}
P_{hh}(k) = b (k)^{2}P_{cdm}(k)\,.\label{bias:1}
\end{eqnarray}
where $b = b_L +1$ is the \textit{Eulerian} halo bias. Equation \eqref{bias:1} is in general not true at all scales in an arbitrary gauge - there are additional terms that arise at near-horizon or larger scales \cite{jeong2012large}. However in the synchronous gauge (which is the gauge we have been working in all along), these additional terms vanish and equation \eqref{bias:1} holds true at \textit{all} scales \cite{jeong2012large}.

Following the results of the previous section, the Lagrangian bias, to a good approximation, is proportional to the response of the linear growth factor $d\log D_{W}/d\delta_{c}$, which depends on $k$ ($=k_{L}$). The scale-dependence of the Eulerian bias with respect to the bias on some reference scale, $k_{\downarrow}$, can then be computed as, 
\begin{eqnarray}
\frac{b(k)}{b(k_{\downarrow})} &=& \left(1-\frac{1}{b}\right)\frac{R_D(k)}{R_D(k_{L\downarrow})}+\frac{1}{b}.
\end{eqnarray}
where here $b$ is the Eulerian bias computed at the scale $k_{\downarrow}$. We choose $k_{\downarrow}$ to be the small scale corresponding to $k_{\downarrow} = 5\times 10^{-2}\ \text{Mpc}^{-1}$. 

We now demonstrate the impact of scale-dependent bias due to radiation for a cosmology with a more realistic radiation density, composed of the usual CMB photon energy density and three massless neutrinos. In Fig. \ref{fig:Pkofk}, we have plotted the scale dependence of the Eulerian halo bias from radiation perturbations for tracers with an Eulerian bias $b= 2$ at small scales. As can be seen from Fig. \ref{fig:Pkofk}, the scale dependence of the halo bias is larger at higher redshifts. The amplitude of this effect depends on the bias of the tracer, as can be seen in Fig. \ref{fig:Pkofb}, where we have plotted the amplitude of the step in the halo bias as a function of the Eulerian bias $b$ at small scales. From both Fig. \ref{fig:Pkofk} and Fig. \ref{fig:Pkofb}, it is evident that the scale dependence of the bias becomes particularly important for the clustering of halos with a large bias, typically $b\geq 2$ at higher redshifts. In particular for objects of bias $b=2$ at small scales, the large-scale bias is less by $0.29\%,\ 0.45\%\ \mathrm{and}\ 0.8\%$ at redshifts $z=0,\ 1\ \mathrm{and}\ 3$ respectively. For $b\gg 1$, these differences approach the corresponding differences in the linear growth factor response; namely $0.43\%,\ 0.68\%\ \mathrm{and}\ 1.2\%$ respectively.

\section{Discussion}

Free-streaming components such as photons and neutrinos cause the evolution of CDM density perturbations to become scale dependent. The free-streaming scale for massless neutrinos or any extra radiation is simply the particle horizon. Subhorizon fluctuations in the energy density of radiation get washed out due to their free streaming, thus suppressing the growth of CDM perturbations on scales smaller than the horizon. This causes the response of small-scale structure formation to large-scale CDM perturbations to become scale dependent. In this paper, we investigated this effect using the Separate Universe technique, wherein the large-scale density perturbation is absorbed into a local expansion history, within a finite subregion of the Universe.

We performed N-body simulations of overdense and underdense Separate Universes with background CDM density perturbations of different scales in the presence of 28 species of massless neutrinos and directly computed the responses of the power spectrum and the halo mass function to the background CDM density perturbations. These responses respectively yield contributions to the squeezed limit CDM bispectrum and the scale-dependent halo bias in a universe with massless neutrinos (extra radiation) along with CMB photons. The large number of massless neutrino species was chosen to produce radiation effects large enough to be reliably measured with a reasonable number of simulations. We ran pairs Separate Universe simulations for two long wavelength perturbations, one near the radiation free-streaming scale at $k_{L} = 5 \times 10^{-4}\ \text{Mpc}^{-1}$, and one below the radiation free-streaming scale at $k_{L} = 5 \times 10 ^{-2}\ \text{Mpc}^{-1}$. 

We found that the scale-dependent responses are insensitive to the composition of the cosmic fluid's radiation component. We tested this by computing the growth responses in the presence of massless neutrinos and photons separately, and found them to be indistinguishable (see Fig. \ref{fig6}). Since the scale-dependent evolution of the long-wavelength matter perturbations is insensitive to the radiation's composition, the halo bias will also be indistinguishable in the presence of photons and massless neutrinos, as long as the total radiation energy density is fixed. 

We detected a non-trivial scale dependence in the responses for the Lagrangian halo bias and the power spectrum. Figures \ref{fig9} and \ref{fig10} show with high statistical significance that the growth contribution to the Separate Universe response of the  power spectrum is dependent on the large $k_L$ at which the Separate Universe is defined. This in turn can be attributed to the scale-dependent growth of the corresponding matter density perturbation $\delta_{c}(k_L, a)$. Figures \ref{fig12} and \ref{fig17} show with high statistical significance that the Separate Universe Lagrangian response bias is dependent on the large-scale wave number, $k_L$, in a similar way.

Figures \ref{fig12} and \ref{fig17} also show that the scale dependence of the Lagrangian halo bias can be accurately characterized by the scale-dependent linear growth response in the Separate Universe, which can be computed numerically using perturbation theory. This is also true of the power spectrum response in the quasilinear regime, as is evident from the agreement between the results of N-body simulations and one-loop perturbation theory calculations of the power spectrum response plotted in Figures \ref{fig9} and \ref{fig10}. 

The agreement between our model and simulation measurements of the scale-dependent Lagrangian response bias is consistent with the assumption that the halo mass function is a universal functional of the CDM power spectrum. Using this fact, we computed the scale dependence of the Eulerian halo bias for the more realistic case of three massless neutrinos and examined its dependence on both the large-scale wave number, and on the small-scale halo bias (Figures \ref{fig:Pkofk} and \ref{fig:Pkofb}). These plots demonstrate that the scale dependence of the Eulerian halo bias is more important for cosmological observations of objects with high bias, and therefore high mass. For this more realistic scenario, the extent of the scale-dependent effects on both the total power spectrum response, and the halo bias, are less than a percent at redshift $z=0$. However, these effects become more important at higher redshifts. Figures \ref{fig4}, \ref{fig:growthresponsemassdep2}, and \ref{Massive_massless_comparison} can be used to estimate the size of the scale-dependent effects for different masses of particles and at different redshifts. 

The scale-dependent changes to the halo bias caused by radiation perturbations are small, yet they are potentially important for future measurements of galaxy clustering on very large scales. Specifically, any measurement that demands accuracy of halo clustering on $\%$-level scales can be affected. A notable example is the search for local-type primordial non-Gaussianity using the scale-dependent bias signature \cite{Dalal:2007cu}. This signature, which exhibits a $\sim 1/k^2$ dependence of the halo bias at horizon scales, is thought to be a particularly robust signature of new physics during the inflationary era. This is because, in a cosmology with Gaussian initial conditions and local structure formation with only CDM and baryons produces {\em constant} bias at large scales. Therefore a detection of a large-scale modulation of the bias can be interpreted as a signature of local primordial non-Gaussianity \cite{Creminelli:2011rh}. Local primordial non-Gaussianity can only be generated in scenarios where additional light degrees of freedom are present during inflation \cite{Creminelli:2004yq}, so its detection would rule out all of the simplest, single-field slow-roll inflation models.  

In this paper, we have demonstrated the existence of a new source of scale-dependent bias on horizon scales that, while having a different $k$-dependence (than the signature of primordial non-Gaussianity), may nevertheless complicate inferences about primordial physics from measurements of large-scale galaxy clustering. While halo formation is still local in our Universe, the local gravitational potential near a halo is now sensitive to both nearby CDM and baryon perturbations as well as radiation perturbations at horizon-scale distances. The gravitational impact of radiation therefore introduces an apparent non-locality to halo bias, similar to the non-gravitational effects of radiation studied in \cite{Sanderbeck:2018lwc}. We leave a detailed study of the impact of scale-dependent bias from radiation on measurements of primordial non-Gaussianity for future work.

\acknowledgements

 Results in this paper were obtained using the high-performance computing system at the Institute for Advanced Computational Science at Stony Brook University. CS received support from DOE DE-SC0017848 and the NASA Grant 80NSSC20K0541. DJ is supported by grant NSF PHY-1620628 and DOE DE-SC0017848. ML is supported by DOE DE-SC0017848.

\newpage

\bibliography{References.bib}

\begin{thebibliography}{41}
\expandafter\ifx\csname natexlab\endcsname\relax\def\natexlab#1{#1}\fi
\expandafter\ifx\csname bibnamefont\endcsname\relax
  \def\bibnamefont#1{#1}\fi
\expandafter\ifx\csname bibfnamefont\endcsname\relax
  \def\bibfnamefont#1{#1}\fi
\expandafter\ifx\csname citenamefont\endcsname\relax
  \def\citenamefont#1{#1}\fi
\expandafter\ifx\csname url\endcsname\relax
  \def\url#1{\texttt{#1}}\fi
\expandafter\ifx\csname urlprefix\endcsname\relax\def\urlprefix{URL }\fi
\providecommand{\bibinfo}[2]{#2}
\providecommand{\eprint}[2][]{\url{#2}}

\bibitem[{\citenamefont{Desjacques et~al.}(2018)\citenamefont{Desjacques,
  Jeong, and Schmidt}}]{Desjacques:2016bnm}
\bibinfo{author}{\bibfnamefont{V.}~\bibnamefont{Desjacques}},
  \bibinfo{author}{\bibfnamefont{D.}~\bibnamefont{Jeong}}, \bibnamefont{and}
  \bibinfo{author}{\bibfnamefont{F.}~\bibnamefont{Schmidt}},
  \bibinfo{journal}{Phys. Rept.} \textbf{\bibinfo{volume}{733}},
  \bibinfo{pages}{1} (\bibinfo{year}{2018}), \eprint{1611.09787}.

\bibitem[{\citenamefont{Bernardeau et~al.}(2002)\citenamefont{Bernardeau,
  Colombi, Gaztanaga, and Scoccimarro}}]{Bernardeau:2001qr}
\bibinfo{author}{\bibfnamefont{F.}~\bibnamefont{Bernardeau}},
  \bibinfo{author}{\bibfnamefont{S.}~\bibnamefont{Colombi}},
  \bibinfo{author}{\bibfnamefont{E.}~\bibnamefont{Gaztanaga}},
  \bibnamefont{and}
  \bibinfo{author}{\bibfnamefont{R.}~\bibnamefont{Scoccimarro}},
  \bibinfo{journal}{Phys. Rept.} \textbf{\bibinfo{volume}{367}},
  \bibinfo{pages}{1} (\bibinfo{year}{2002}), \eprint{astro-ph/0112551}.

\bibitem[{\citenamefont{Lesgourgues and Pastor}(2006)}]{lesgourgues2006massive}
\bibinfo{author}{\bibfnamefont{J.}~\bibnamefont{Lesgourgues}} \bibnamefont{and}
  \bibinfo{author}{\bibfnamefont{S.}~\bibnamefont{Pastor}},
  \bibinfo{journal}{Physics Reports} \textbf{\bibinfo{volume}{429}},
  \bibinfo{pages}{307} (\bibinfo{year}{2006}).

\bibitem[{\citenamefont{Dor\'e et~al.}(2014)}]{Dore:2014cca}
\bibinfo{author}{\bibfnamefont{O.}~\bibnamefont{Dor\'e}} \bibnamefont{et~al.}
  (\bibinfo{year}{2014}), \eprint{1412.4872}.

\bibitem[{\citenamefont{Meerburg et~al.}(2019)}]{Meerburg:2019qqi}
\bibinfo{author}{\bibfnamefont{P.~D.} \bibnamefont{Meerburg}}
  \bibnamefont{et~al.} (\bibinfo{year}{2019}), \eprint{1903.04409}.

\bibitem[{\citenamefont{McDonald}(2003)}]{McDonald:2001fe}
\bibinfo{author}{\bibfnamefont{P.}~\bibnamefont{McDonald}},
  \bibinfo{journal}{Astrophys. J.} \textbf{\bibinfo{volume}{585}},
  \bibinfo{pages}{34} (\bibinfo{year}{2003}), \eprint{astro-ph/0108064}.

\bibitem[{\citenamefont{Sirko}(2005)}]{Sirko:2005uz}
\bibinfo{author}{\bibfnamefont{E.}~\bibnamefont{Sirko}},
  \bibinfo{journal}{Astrophys. J.} \textbf{\bibinfo{volume}{634}},
  \bibinfo{pages}{728} (\bibinfo{year}{2005}), \eprint{astro-ph/0503106}.

\bibitem[{\citenamefont{Gnedin et~al.}(2011)\citenamefont{Gnedin, Kravtsov, and
  Rudd}}]{Gnedin:2011kj}
\bibinfo{author}{\bibfnamefont{N.~Y.} \bibnamefont{Gnedin}},
  \bibinfo{author}{\bibfnamefont{A.~V.} \bibnamefont{Kravtsov}},
  \bibnamefont{and} \bibinfo{author}{\bibfnamefont{D.~H.} \bibnamefont{Rudd}},
  \bibinfo{journal}{Astrophys. J. Suppl.} \textbf{\bibinfo{volume}{194}},
  \bibinfo{pages}{46} (\bibinfo{year}{2011}), \eprint{1104.1428}.

\bibitem[{\citenamefont{Wagner et~al.}(2015)\citenamefont{Wagner, Schmidt,
  Chiang, and Komatsu}}]{Wagner:2014aka}
\bibinfo{author}{\bibfnamefont{C.}~\bibnamefont{Wagner}},
  \bibinfo{author}{\bibfnamefont{F.}~\bibnamefont{Schmidt}},
  \bibinfo{author}{\bibfnamefont{C.-T.} \bibnamefont{Chiang}},
  \bibnamefont{and} \bibinfo{author}{\bibfnamefont{E.}~\bibnamefont{Komatsu}},
  \bibinfo{journal}{Mon. Not. Roy. Astron. Soc.}
  \textbf{\bibinfo{volume}{448}}, \bibinfo{pages}{L11} (\bibinfo{year}{2015}),
  \eprint{1409.6294}.

\bibitem[{\citenamefont{Li et~al.}(2014{\natexlab{a}})\citenamefont{Li, Hu, and
  Takada}}]{Li:2014sga}
\bibinfo{author}{\bibfnamefont{Y.}~\bibnamefont{Li}},
  \bibinfo{author}{\bibfnamefont{W.}~\bibnamefont{Hu}}, \bibnamefont{and}
  \bibinfo{author}{\bibfnamefont{M.}~\bibnamefont{Takada}},
  \bibinfo{journal}{Phys. Rev. D} \textbf{\bibinfo{volume}{89}},
  \bibinfo{pages}{083519} (\bibinfo{year}{2014}{\natexlab{a}}),
  \eprint{1401.0385}.

\bibitem[{\citenamefont{Li et~al.}(2014{\natexlab{b}})\citenamefont{Li, Hu, and
  Takada}}]{Li:2014jra}
\bibinfo{author}{\bibfnamefont{Y.}~\bibnamefont{Li}},
  \bibinfo{author}{\bibfnamefont{W.}~\bibnamefont{Hu}}, \bibnamefont{and}
  \bibinfo{author}{\bibfnamefont{M.}~\bibnamefont{Takada}},
  \bibinfo{journal}{Phys. Rev.} \textbf{\bibinfo{volume}{D90}},
  \bibinfo{pages}{103530} (\bibinfo{year}{2014}{\natexlab{b}}),
  \eprint{1408.1081}.

\bibitem[{\citenamefont{Li et~al.}(2016{\natexlab{a}})\citenamefont{Li, Hu, and
  Takada}}]{Li:2015jsz}
\bibinfo{author}{\bibfnamefont{Y.}~\bibnamefont{Li}},
  \bibinfo{author}{\bibfnamefont{W.}~\bibnamefont{Hu}}, \bibnamefont{and}
  \bibinfo{author}{\bibfnamefont{M.}~\bibnamefont{Takada}},
  \bibinfo{journal}{Phys. Rev.} \textbf{\bibinfo{volume}{D93}},
  \bibinfo{pages}{063507} (\bibinfo{year}{2016}{\natexlab{a}}),
  \eprint{1511.01454}.

\bibitem[{\citenamefont{Chiang et~al.}(2014)\citenamefont{Chiang, Wagner,
  Schmidt, and Komatsu}}]{Chiang:2014oga}
\bibinfo{author}{\bibfnamefont{C.-T.} \bibnamefont{Chiang}},
  \bibinfo{author}{\bibfnamefont{C.}~\bibnamefont{Wagner}},
  \bibinfo{author}{\bibfnamefont{F.}~\bibnamefont{Schmidt}}, \bibnamefont{and}
  \bibinfo{author}{\bibfnamefont{E.}~\bibnamefont{Komatsu}},
  \bibinfo{journal}{JCAP} \textbf{\bibinfo{volume}{1405}}, \bibinfo{pages}{048}
  (\bibinfo{year}{2014}), \eprint{1403.3411}.

\bibitem[{\citenamefont{Manzotti et~al.}(2014)\citenamefont{Manzotti, Hu, and
  Benoit-Lévy}}]{Manzotti:2014wca}
\bibinfo{author}{\bibfnamefont{A.}~\bibnamefont{Manzotti}},
  \bibinfo{author}{\bibfnamefont{W.}~\bibnamefont{Hu}}, \bibnamefont{and}
  \bibinfo{author}{\bibfnamefont{A.}~\bibnamefont{Benoit-Lévy}},
  \bibinfo{journal}{Phys. Rev.} \textbf{\bibinfo{volume}{D90}},
  \bibinfo{pages}{023003} (\bibinfo{year}{2014}), \eprint{1401.7992}.

\bibitem[{\citenamefont{Baldauf et~al.}(2016)\citenamefont{Baldauf, Seljak,
  Senatore, and Zaldarriaga}}]{Baldauf:2015vio}
\bibinfo{author}{\bibfnamefont{T.}~\bibnamefont{Baldauf}},
  \bibinfo{author}{\bibfnamefont{U.}~\bibnamefont{Seljak}},
  \bibinfo{author}{\bibfnamefont{L.}~\bibnamefont{Senatore}}, \bibnamefont{and}
  \bibinfo{author}{\bibfnamefont{M.}~\bibnamefont{Zaldarriaga}},
  \bibinfo{journal}{JCAP} \textbf{\bibinfo{volume}{1609}}, \bibinfo{pages}{007}
  (\bibinfo{year}{2016}), \eprint{1511.01465}.

\bibitem[{\citenamefont{Lazeyras et~al.}(2016)\citenamefont{Lazeyras, Wagner,
  Baldauf, and Schmidt}}]{Lazeyras:2015lgp}
\bibinfo{author}{\bibfnamefont{T.}~\bibnamefont{Lazeyras}},
  \bibinfo{author}{\bibfnamefont{C.}~\bibnamefont{Wagner}},
  \bibinfo{author}{\bibfnamefont{T.}~\bibnamefont{Baldauf}}, \bibnamefont{and}
  \bibinfo{author}{\bibfnamefont{F.}~\bibnamefont{Schmidt}},
  \bibinfo{journal}{JCAP} \textbf{\bibinfo{volume}{1602}}, \bibinfo{pages}{018}
  (\bibinfo{year}{2016}), \eprint{1511.01096}.

\bibitem[{\citenamefont{Paranjape and Padmanabhan}(2017)}]{Paranjape:2016pbh}
\bibinfo{author}{\bibfnamefont{A.}~\bibnamefont{Paranjape}} \bibnamefont{and}
  \bibinfo{author}{\bibfnamefont{N.}~\bibnamefont{Padmanabhan}},
  \bibinfo{journal}{Mon. Not. Roy. Astron. Soc.}
  \textbf{\bibinfo{volume}{468}}, \bibinfo{pages}{2984} (\bibinfo{year}{2017}),
  \eprint{1612.02833}.

\bibitem[{\citenamefont{Chan et~al.}(2019)\citenamefont{Chan, Li, Biagetti, and
  Hamaus}}]{Chan:2019yzq}
\bibinfo{author}{\bibfnamefont{K.~C.} \bibnamefont{Chan}},
  \bibinfo{author}{\bibfnamefont{Y.}~\bibnamefont{Li}},
  \bibinfo{author}{\bibfnamefont{M.}~\bibnamefont{Biagetti}}, \bibnamefont{and}
  \bibinfo{author}{\bibfnamefont{N.}~\bibnamefont{Hamaus}}
  (\bibinfo{year}{2019}), \eprint{1909.03736}.

\bibitem[{\citenamefont{Jamieson and Loverde}(2019)}]{Jamieson:2019dmp}
\bibinfo{author}{\bibfnamefont{D.}~\bibnamefont{Jamieson}} \bibnamefont{and}
  \bibinfo{author}{\bibfnamefont{M.}~\bibnamefont{Loverde}},
  \bibinfo{journal}{Phys. Rev. D} \textbf{\bibinfo{volume}{100}},
  \bibinfo{pages}{123528} (\bibinfo{year}{2019}), \eprint{1909.05313}.

\bibitem[{\citenamefont{Jamieson and Loverde}(2020)}]{Jamieson:2020wxf}
\bibinfo{author}{\bibfnamefont{D.}~\bibnamefont{Jamieson}} \bibnamefont{and}
  \bibinfo{author}{\bibfnamefont{M.}~\bibnamefont{Loverde}}
  (\bibinfo{year}{2020}), \eprint{2010.07235}.

\bibitem[{\citenamefont{Hu et~al.}(2016)\citenamefont{Hu, Chiang, Li, and
  LoVerde}}]{Hu:2016ssz}
\bibinfo{author}{\bibfnamefont{W.}~\bibnamefont{Hu}},
  \bibinfo{author}{\bibfnamefont{C.-T.} \bibnamefont{Chiang}},
  \bibinfo{author}{\bibfnamefont{Y.}~\bibnamefont{Li}}, \bibnamefont{and}
  \bibinfo{author}{\bibfnamefont{M.}~\bibnamefont{LoVerde}},
  \bibinfo{journal}{Phys. Rev.} \textbf{\bibinfo{volume}{D94}},
  \bibinfo{pages}{023002} (\bibinfo{year}{2016}), \eprint{1605.01412}.

\bibitem[{\citenamefont{Chiang et~al.}(2017)\citenamefont{Chiang, Hu, Li, and
  Loverde}}]{Chiang:2017vuk}
\bibinfo{author}{\bibfnamefont{C.-T.} \bibnamefont{Chiang}},
  \bibinfo{author}{\bibfnamefont{W.}~\bibnamefont{Hu}},
  \bibinfo{author}{\bibfnamefont{Y.}~\bibnamefont{Li}}, \bibnamefont{and}
  \bibinfo{author}{\bibfnamefont{M.}~\bibnamefont{Loverde}}
  (\bibinfo{year}{2017}), \eprint{1710.01310}.

\bibitem[{\citenamefont{Chiang et~al.}(2016)\citenamefont{Chiang, Li, Hu, and
  LoVerde}}]{Chiang:2016vxa}
\bibinfo{author}{\bibfnamefont{C.-T.} \bibnamefont{Chiang}},
  \bibinfo{author}{\bibfnamefont{Y.}~\bibnamefont{Li}},
  \bibinfo{author}{\bibfnamefont{W.}~\bibnamefont{Hu}}, \bibnamefont{and}
  \bibinfo{author}{\bibfnamefont{M.}~\bibnamefont{LoVerde}},
  \bibinfo{journal}{Phys. Rev.} \textbf{\bibinfo{volume}{D94}},
  \bibinfo{pages}{123502} (\bibinfo{year}{2016}), \eprint{1609.01701}.

\bibitem[{\citenamefont{Jamieson and LoVerde}(2019)}]{Jamieson:2018biz}
\bibinfo{author}{\bibfnamefont{D.}~\bibnamefont{Jamieson}} \bibnamefont{and}
  \bibinfo{author}{\bibfnamefont{M.}~\bibnamefont{LoVerde}},
  \bibinfo{journal}{Phys. Rev.} \textbf{\bibinfo{volume}{D100}},
  \bibinfo{pages}{023516} (\bibinfo{year}{2019}), \eprint{1812.08765}.

\bibitem[{\citenamefont{Barreira et~al.}(2020)\citenamefont{Barreira, Cabass,
  Nelson, and Schmidt}}]{Barreira:2019qdl}
\bibinfo{author}{\bibfnamefont{A.}~\bibnamefont{Barreira}},
  \bibinfo{author}{\bibfnamefont{G.}~\bibnamefont{Cabass}},
  \bibinfo{author}{\bibfnamefont{D.}~\bibnamefont{Nelson}}, \bibnamefont{and}
  \bibinfo{author}{\bibfnamefont{F.}~\bibnamefont{Schmidt}},
  \bibinfo{journal}{JCAP} \textbf{\bibinfo{volume}{02}}, \bibinfo{pages}{005}
  (\bibinfo{year}{2020}), \eprint{1907.04317}.

\bibitem[{\citenamefont{Pisanti et~al.}(2008)\citenamefont{Pisanti, Cirillo,
  Esposito, Iocco, Mangano, Miele, and Serpico}}]{Pisanti:2007hk}
\bibinfo{author}{\bibfnamefont{O.}~\bibnamefont{Pisanti}},
  \bibinfo{author}{\bibfnamefont{A.}~\bibnamefont{Cirillo}},
  \bibinfo{author}{\bibfnamefont{S.}~\bibnamefont{Esposito}},
  \bibinfo{author}{\bibfnamefont{F.}~\bibnamefont{Iocco}},
  \bibinfo{author}{\bibfnamefont{G.}~\bibnamefont{Mangano}},
  \bibinfo{author}{\bibfnamefont{G.}~\bibnamefont{Miele}}, \bibnamefont{and}
  \bibinfo{author}{\bibfnamefont{P.}~\bibnamefont{Serpico}},
  \bibinfo{journal}{Comput. Phys. Commun.} \textbf{\bibinfo{volume}{178}},
  \bibinfo{pages}{956} (\bibinfo{year}{2008}), \eprint{0705.0290}.

\bibitem[{\citenamefont{Blas et~al.}(2011)\citenamefont{Blas, Lesgourgues, and
  Tram}}]{blas2011cosmic}
\bibinfo{author}{\bibfnamefont{D.}~\bibnamefont{Blas}},
  \bibinfo{author}{\bibfnamefont{J.}~\bibnamefont{Lesgourgues}},
  \bibnamefont{and} \bibinfo{author}{\bibfnamefont{T.}~\bibnamefont{Tram}},
  \bibinfo{journal}{Journal of Cosmology and Astroparticle Physics}
  \textbf{\bibinfo{volume}{2011}}, \bibinfo{pages}{034} (\bibinfo{year}{2011}).

\bibitem[{\citenamefont{Springel}(2005)}]{springel2005cosmological}
\bibinfo{author}{\bibfnamefont{V.}~\bibnamefont{Springel}},
  \bibinfo{journal}{Monthly notices of the royal astronomical society}
  \textbf{\bibinfo{volume}{364}}, \bibinfo{pages}{1105} (\bibinfo{year}{2005}).

\bibitem[{\citenamefont{Crocce et~al.}(2006)\citenamefont{Crocce, Pueblas, and
  Scoccimarro}}]{Crocce:2006ve}
\bibinfo{author}{\bibfnamefont{M.}~\bibnamefont{Crocce}},
  \bibinfo{author}{\bibfnamefont{S.}~\bibnamefont{Pueblas}}, \bibnamefont{and}
  \bibinfo{author}{\bibfnamefont{R.}~\bibnamefont{Scoccimarro}},
  \bibinfo{journal}{Mon. Not. Roy. Astron. Soc.}
  \textbf{\bibinfo{volume}{373}}, \bibinfo{pages}{369} (\bibinfo{year}{2006}),
  \eprint{astro-ph/0606505}.

\bibitem[{\citenamefont{Chiang et~al.}(2018)\citenamefont{Chiang, Hu, Li, and
  LoVerde}}]{chiang2018scale}
\bibinfo{author}{\bibfnamefont{C.-T.} \bibnamefont{Chiang}},
  \bibinfo{author}{\bibfnamefont{W.}~\bibnamefont{Hu}},
  \bibinfo{author}{\bibfnamefont{Y.}~\bibnamefont{Li}}, \bibnamefont{and}
  \bibinfo{author}{\bibfnamefont{M.}~\bibnamefont{LoVerde}},
  \bibinfo{journal}{Physical Review D} \textbf{\bibinfo{volume}{97}},
  \bibinfo{pages}{123526} (\bibinfo{year}{2018}).

\bibitem[{\citenamefont{Behroozi et~al.}(2012)\citenamefont{Behroozi, Wechsler,
  and Wu}}]{behroozi2012rockstar}
\bibinfo{author}{\bibfnamefont{P.~S.} \bibnamefont{Behroozi}},
  \bibinfo{author}{\bibfnamefont{R.~H.} \bibnamefont{Wechsler}},
  \bibnamefont{and} \bibinfo{author}{\bibfnamefont{H.-Y.} \bibnamefont{Wu}},
  \bibinfo{journal}{The Astrophysical Journal} \textbf{\bibinfo{volume}{762}},
  \bibinfo{pages}{109} (\bibinfo{year}{2012}).

\bibitem[{\citenamefont{Li et~al.}(2016{\natexlab{b}})\citenamefont{Li, Hu, and
  Takada}}]{li2016separate}
\bibinfo{author}{\bibfnamefont{Y.}~\bibnamefont{Li}},
  \bibinfo{author}{\bibfnamefont{W.}~\bibnamefont{Hu}}, \bibnamefont{and}
  \bibinfo{author}{\bibfnamefont{M.}~\bibnamefont{Takada}},
  \bibinfo{journal}{Physical Review D} \textbf{\bibinfo{volume}{93}},
  \bibinfo{pages}{063507} (\bibinfo{year}{2016}{\natexlab{b}}).

\bibitem[{\citenamefont{Frigo and Johnson}(2005)}]{frigo2005design}
\bibinfo{author}{\bibfnamefont{M.}~\bibnamefont{Frigo}} \bibnamefont{and}
  \bibinfo{author}{\bibfnamefont{S.~G.} \bibnamefont{Johnson}},
  \bibinfo{journal}{Proceedings of the IEEE} \textbf{\bibinfo{volume}{93}},
  \bibinfo{pages}{216} (\bibinfo{year}{2005}).

\bibitem[{\citenamefont{Parfrey et~al.}(2011)\citenamefont{Parfrey, Hui, and
  Sheth}}]{Parfrey:2010uy}
\bibinfo{author}{\bibfnamefont{K.}~\bibnamefont{Parfrey}},
  \bibinfo{author}{\bibfnamefont{L.}~\bibnamefont{Hui}}, \bibnamefont{and}
  \bibinfo{author}{\bibfnamefont{R.~K.} \bibnamefont{Sheth}},
  \bibinfo{journal}{Phys. Rev. D} \textbf{\bibinfo{volume}{83}},
  \bibinfo{pages}{063511} (\bibinfo{year}{2011}), \eprint{1012.1335}.

\bibitem[{\citenamefont{LoVerde}(2014)}]{LoVerde:2014pxa}
\bibinfo{author}{\bibfnamefont{M.}~\bibnamefont{LoVerde}},
  \bibinfo{journal}{Phys. Rev. D} \textbf{\bibinfo{volume}{90}},
  \bibinfo{pages}{083530} (\bibinfo{year}{2014}), \eprint{1405.4855}.

\bibitem[{\citenamefont{Chiang et~al.}(2019)\citenamefont{Chiang, LoVerde, and
  Villaescusa-Navarro}}]{Chiang:2018laa}
\bibinfo{author}{\bibfnamefont{C.-T.} \bibnamefont{Chiang}},
  \bibinfo{author}{\bibfnamefont{M.}~\bibnamefont{LoVerde}}, \bibnamefont{and}
  \bibinfo{author}{\bibfnamefont{F.}~\bibnamefont{Villaescusa-Navarro}},
  \bibinfo{journal}{Phys. Rev. Lett.} \textbf{\bibinfo{volume}{122}},
  \bibinfo{pages}{041302} (\bibinfo{year}{2019}), \eprint{1811.12412}.

\bibitem[{\citenamefont{Jeong et~al.}(2012)\citenamefont{Jeong, Schmidt, and
  Hirata}}]{jeong2012large}
\bibinfo{author}{\bibfnamefont{D.}~\bibnamefont{Jeong}},
  \bibinfo{author}{\bibfnamefont{F.}~\bibnamefont{Schmidt}}, \bibnamefont{and}
  \bibinfo{author}{\bibfnamefont{C.~M.} \bibnamefont{Hirata}},
  \bibinfo{journal}{Physical Review D} \textbf{\bibinfo{volume}{85}},
  \bibinfo{pages}{023504} (\bibinfo{year}{2012}).

\bibitem[{\citenamefont{Dalal et~al.}(2008)\citenamefont{Dalal, Dore, Huterer,
  and Shirokov}}]{Dalal:2007cu}
\bibinfo{author}{\bibfnamefont{N.}~\bibnamefont{Dalal}},
  \bibinfo{author}{\bibfnamefont{O.}~\bibnamefont{Dore}},
  \bibinfo{author}{\bibfnamefont{D.}~\bibnamefont{Huterer}}, \bibnamefont{and}
  \bibinfo{author}{\bibfnamefont{A.}~\bibnamefont{Shirokov}},
  \bibinfo{journal}{Phys. Rev. D} \textbf{\bibinfo{volume}{77}},
  \bibinfo{pages}{123514} (\bibinfo{year}{2008}), \eprint{0710.4560}.

\bibitem[{\citenamefont{Creminelli et~al.}(2011)\citenamefont{Creminelli,
  D'Amico, Musso, and Norena}}]{Creminelli:2011rh}
\bibinfo{author}{\bibfnamefont{P.}~\bibnamefont{Creminelli}},
  \bibinfo{author}{\bibfnamefont{G.}~\bibnamefont{D'Amico}},
  \bibinfo{author}{\bibfnamefont{M.}~\bibnamefont{Musso}}, \bibnamefont{and}
  \bibinfo{author}{\bibfnamefont{J.}~\bibnamefont{Norena}},
  \bibinfo{journal}{JCAP} \textbf{\bibinfo{volume}{11}}, \bibinfo{pages}{038}
  (\bibinfo{year}{2011}), \eprint{1106.1462}.

\bibitem[{\citenamefont{Creminelli and Zaldarriaga}(2004)}]{Creminelli:2004yq}
\bibinfo{author}{\bibfnamefont{P.}~\bibnamefont{Creminelli}} \bibnamefont{and}
  \bibinfo{author}{\bibfnamefont{M.}~\bibnamefont{Zaldarriaga}},
  \bibinfo{journal}{JCAP} \textbf{\bibinfo{volume}{10}}, \bibinfo{pages}{006}
  (\bibinfo{year}{2004}), \eprint{astro-ph/0407059}.

\bibitem[{\citenamefont{Sanderbeck et~al.}(2019)\citenamefont{Sanderbeck,
  Ir\v{s}i\v{c}, McQuinn, and Meiksin}}]{Sanderbeck:2018lwc}
\bibinfo{author}{\bibfnamefont{P.~U.} \bibnamefont{Sanderbeck}},
  \bibinfo{author}{\bibfnamefont{V.}~\bibnamefont{Ir\v{s}i\v{c}}},
  \bibinfo{author}{\bibfnamefont{M.}~\bibnamefont{McQuinn}}, \bibnamefont{and}
  \bibinfo{author}{\bibfnamefont{A.}~\bibnamefont{Meiksin}},
  \bibinfo{journal}{Mon. Not. Roy. Astron. Soc.}
  \textbf{\bibinfo{volume}{485}}, \bibinfo{pages}{5059} (\bibinfo{year}{2019}),
  \eprint{1810.12321}.

\end{thebibliography}

\pagenumbering{gobble}

\end{document}